# Towards a Mathematical Theory of the Delays of the Asynchronous Circuits

Serban E. Vlad
str. Zimbrului, Nr. 3, Bl. PB68, Et. 2, Ap. 11, 3700, Oradea, Romania
http://site.voila.fr/serban_e_vlad, email: serban_e_vlad@yahoo.com

**Abstract** *The inequations of the delays of the asynchronous circuits are written, by making use of pseudo-Boolean differential calculus. We consider these efforts to be a possible starting point in the semi-formalized reconstruction of the digital electrical engineering* (*which is a non-formalized theory*).

**List of Abbreviations**

BDC - bounded delay condition, in 4.1
BDE - bounded delay element, in 4.12
BRIDC - bounded (bivalent) relative inertial delay condition, in 7.1
BRIDE - bounded (bivalent) relative inertial delay element, in 7.11
CC - consistency condition, in 4.4 and in 7.5
DC - delay condition, in 3.2
DE - delay element, in 3.8
DRIDC - determinisic (bivalent) relative inertial delay condition, in 8.6
DRIDE - deterministic (bivalent) relative inertial delay element, in 8.12
FDC - fixed delay condition, in 5.1
FDE - fixed delay element, in 5.8
NZC - non-zenoness conditions, in 6.9
RIDC - (bivalent) relative inertial delay condition, in 6.2
RIDE - (bivalent) relative inertial delay element, in 6.13
SC - stability condition, in 3.1

**Contents**

## 1. Introduction

Digital electrical engineering[1] is in present a non-formalized theory, together with these chapters from computer science that treat topics of electronics. Our purpose is that of looking for and proposing concepts and a language that are suitable for the reconstruction of digital electrical engineering as a semi-formalized theory. Our belief is that the concepts of *Boolean functions* and *delay elements* and the language of the (*differential*) *equations and inequations on pseudo-Boolean functions* (i.e. the $\mathbf{R} \to \{0,1\}$ functions) satisfy this request.

As revealed in this moment, the main problems and questions arising are:

- *are our definitions of the delay elements acceptable ?*
- *how accessible is such a theory ?*
- *how easy is applying the theory in problems of analysis, synthesis and verification of the asynchronous circuits ?*

We do not think of easy or fast steps. In fact, the situation of crisis recalls the one from mathematics that was generated by the naive sets theory of Cantor, where capital shortcomings were found: the well known paradoxes. In our work [11] we have found:

- the paradox that the notion of *inertial delay buffer* was incorrectly defined in the literature (by some important authors), i.e. against the intuition (accepted by the same authors) and a simple counterexample showed this fact
- the apparent paradox that this elementary notion, after being

[1] including the asynchronous automata, timed automata, asynchronous circuits, Boolean circuits, switching circuits, asynchronous systems etc. all these refer to more or less the same topic treated perhaps differently by different authors

incorrectly defined, did not stop or (strongly) influence the exposure and this is a feature of the non-formalized theories

- the paradox that, after a thorough definition of the inertial delay buffer closely related with the (generally accepted) intuition, the concept did not offer the expected property of closure: the serial connection of the inertial delay buffers is not an inertial delay buffer.

Thus, serious reasons existed for increasing the efforts of finding a *good start* in digital electrical engineering.

Basically, the delays, respectively the delay elements[2] are related with the computation of the identity function on {0,1} and they are the most simple circuits from electronics. A common sense classification is the following one:

**I** Delays on gates, on wires and combined delays, on gates and wires. We are not interested in this approach, our models may be applied where necessary

**II** a) <u>unbounded delays</u>, *'if no bound on the magnitude is known a priori, except that it is positive and finite'*, [2].

b) <u>bounded delays</u>, a delay is bounded *'if an upper and lower bound on its magnitude are known before the synthesis or analysis of the circuit is begun'*, [2]. In [3], the definition and the explanation are the following: *'each component is assumed to have an uncertain delay that lies between given upper and lower bounds. The delay bounds take into account potential delay variations due to statistical fluctuations in the fabrication process, variations in ambient temperature, power supply, etc'*.

c) <u>fixed delays</u>, special case of bounded delays, when the upper and the lower bounds of the delays coincide and the uncertainties characterizing the delays disappear; thus the delays are known.

**III** a) <u>pure delays;</u> such a delay is defined in [2] by: *'it transmits each event on its input to its output, i.e. it corresponds to a pure translation in time of the input waveform"*.

b) <u>inertial delays</u>; we refer to [2] once again: *'pulses shorter than or equal to the delay magnitude are not transmitted'*. This concept is a simplification however since we have two parameters here: one making the pulses shorter than or equal to it be not transmitted (called *cancellation*

[2] the delays are real numbers and the delay elements are either circuits, or theoretical abstractions of these circuits that are used in modeling. Sometimes, the word 'delay' is used as a short form for any of these meanings.

*delay* in [1]) and the other representing the delay with which the input is transmitted to the output (called *transmission delay for transitions* in [1]).

We shall present now the informal definition of the inertial delay buffer that is, in a certain sense, the key definition of our work. By *informal definition* we mean presenting the behaviour of the circuit.

In Fig 1, the pairs $(u(t),x(t))$ represent the input and the output of the inertial delay buffer at a certain time instant and $a,b,c,d$ are the labels of the transitions from one pair input-output to another pair. We suppose that the initial position of the circuit is (0,0) and this position is of equilibrium, meaning that the circuit can remain indefinitely long in it if the input remains constant 0 indefinitely long. The other position of equilibrium is (1,1), the circuit can remain in it indefinitely long if the input remains constant 1 indefinitely long. The parameters $0 < d_{r,\min} \leq d_{r,\max}$, $0 < d_{f,\min} \leq d_{f,\max}$ are given.

*a*: We suppose that at the time instant $t_1$ the input switches from 0 to 1

*b*: Let's say that the input is constant 1 from $t_1$ until it switches at

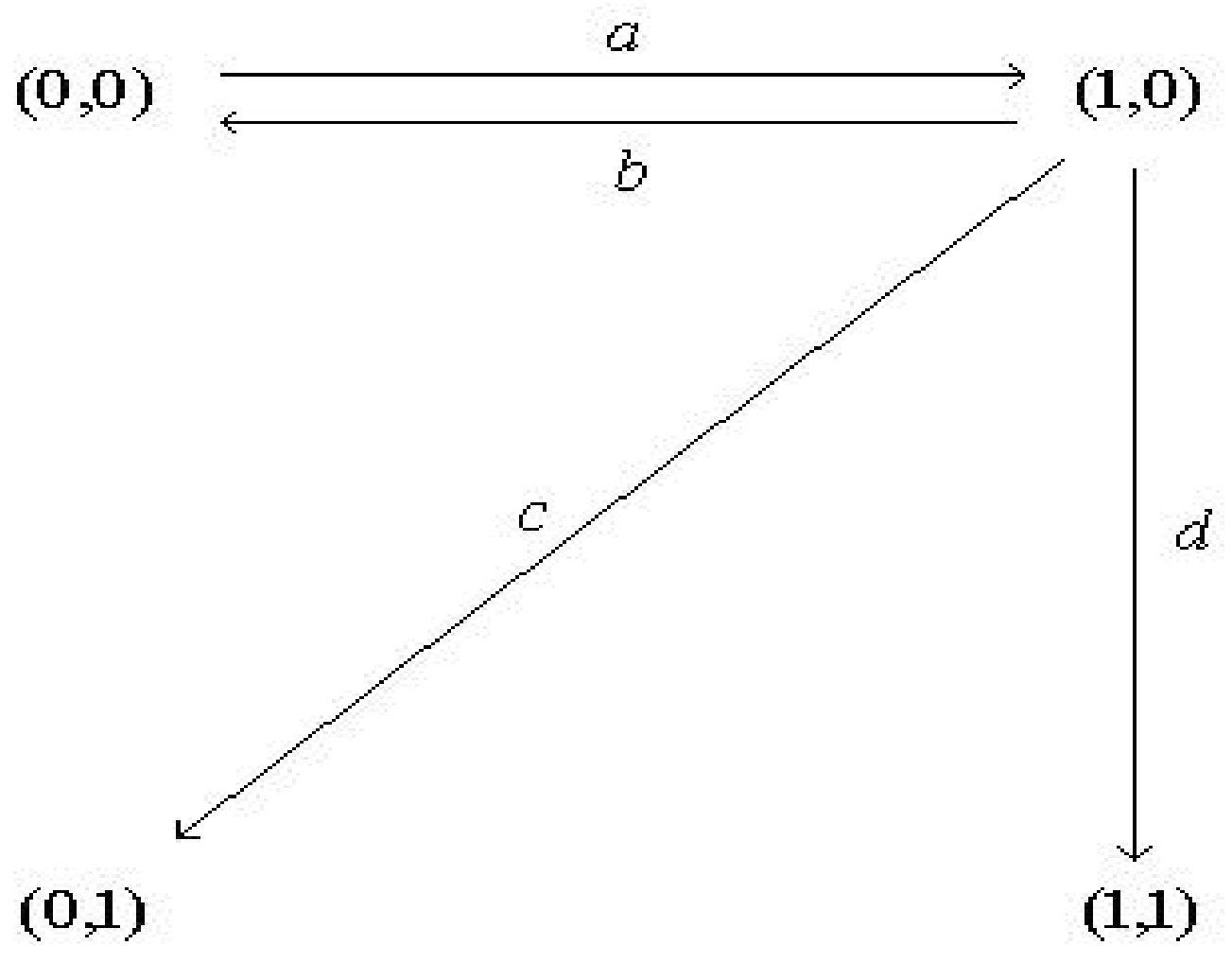

Fig 1

the time instant $t_2 \in (t_1, t_1 + d_{r,\max})$ from 1 back to 0. If $t_2 \in (t_1, t_1 + d_{r,\min})$ then $b$ is surely run and if $t_2 \in [t_1 + d_{r,\min}, t_1 + d_{r,\max})$, then running $b$ is possible

*c*: if the input is constant 1 from $t_1$ until it switches at the time instant $t_2 \in [t_1 + d_{r,\min}, t_1 + d_{r,\max})$ from 1 back to 0, then running $c$ is possible too

*d*: Let's suppose that the input remains constant 1 from $t_1$ until the time instant $t_2 \geq t_1 + d_{r,\min}$. If $t_2 \in [t_1 + d_{r,\min}, t_1 + d_{r,\max})$, running $d$ is possible and if $t_2 > t_1 + d_{r,\max}$, then $d$ was already run.

Replacing 1 with 0, 0 with 1 and 'r' with 'f' gives the dual behaviour of the circuit.

The paper is structured in chapters and each chapter has several paragraphs. The most important equations, inequations and logical conditions are numbered with (1), (2),… inside each paragraph. Lists exist also: a), b),… ,i), ii),… inside some paragraphs. When referring to them, 2.7 (3), 2.7 b) mean for example item (3), respectively item b) from chapter 2, paragraph 7. The tables and the figures are numbered continuously: 1, 2, 3,…

The content of the paper is the following. In chapter 2 we present some important notions of pseudo-Boolean calculus making the paper self-contained. In chapter 3 we have the definition of the delays. Chapters 4,5,6,7,8 present the bounded delays, the fixed delays, the inertial delays, the bounded inertial delays and the deterministic inertial delays. Chapter 9 makes a comparison between the points of view expressed here and in some previous works of the author. Chapter 10 is of conclusions.

## 2. Preliminaries

2.1 **Definition** $\boldsymbol{B} = \{0,1\}$ is endowed with the *discrete topology*, where the open sets are all the subsets, with the order $0 \leq 1$ and with the usual laws: the *complement* $'\overline{\ }'$, the *reunion* $'\cup'$, the *product* $'\cdot'$, the *modulo 2 sum* $'\oplus'$ etc:

| $-$ | 0 | 1 |
|---|---|---|
| | 1 | 0 |

a)

| $\cup$ | 0 | 1 |
|---|---|---|
| 0 | 0 | 1 |
| 1 | 1 | 1 |

b)

| $\cdot$ | 0 | 1 |
|---|---|---|
| 0 | 0 | 0 |
| 1 | 0 | 1 |

c)

| $\oplus$ | 0 | 1 |
|---|---|---|
| 0 | 0 | 1 |
| 1 | 1 | 0 |

d)

Table 1

2.2 **Notation** $\mathbf{0},\mathbf{1}:\boldsymbol{R}\to\boldsymbol{B}$ are the two constant functions.

2.3 **Notations** In the set $2^{\boldsymbol{R}}$ of the subsets of $\boldsymbol{R}$, we note with '-' the set difference, with $'\vee'$ the set reunion, with $'\wedge'$ the set intersection and with $'\Delta'$ the set symmetrical difference. We keep the notations $'\,\overline{\ }\,','\cup','\cdot','\oplus'$ for the laws that are induced by $\boldsymbol{B}$ on the set of the functions $\boldsymbol{R}\to\boldsymbol{B}$.

2.4 **Definition** $\chi_A:\boldsymbol{R}\to\boldsymbol{B}$ is the *characteristic function* of the set $A\subset\boldsymbol{R}$:

$$\chi_A(t)=\begin{cases}1, t\in A\\ 0, t\notin A\end{cases} \qquad (1)$$

2.5 **Definition** The *support (set) of the function* $w:\boldsymbol{R}\to\boldsymbol{B}$ is the set $supp\ w\subset\boldsymbol{R}$ defined in the next manner:

$$supp\ w=\{t\mid t\in\boldsymbol{R}, w(t)=1\} \qquad (1)$$

2.6 **Theorem** We have

$$\chi_\emptyset=\mathbf{0} \qquad supp\,\mathbf{0}=\emptyset \qquad (1)$$

$$\chi_{\boldsymbol{R}}=\mathbf{1} \qquad supp\,\mathbf{1}=\boldsymbol{R} \qquad (2)$$

$$\forall t, w(t)=\chi_{supp\ w}(t) \qquad supp\ w=supp\ \chi_{supp\ w} \qquad (3)$$

$$\forall t, \overline{w(t)}=\chi_{\boldsymbol{R}-supp\ w}(t) \qquad supp\ \overline{w}=\boldsymbol{R}-supp\ w \qquad (4)$$

$$\forall t, w(t)\cup w'(t)=\chi_{supp\ w\vee supp\ w'}(t)$$

$$supp\,(w\cup w')=supp\ w\vee supp\ w' \qquad (5)$$

$$\forall t, w(t)\cdot w'(t)=\chi_{supp\ w\wedge supp\ w'}(t)$$

$$supp\,(w\cdot w')=supp\ w\wedge supp\ w' \qquad (6)$$

$$\forall t, w(t)\oplus w'(t)=\chi_{supp\ w\Delta\ supp\ w'}(t)$$

$$supp\,(w\oplus w')=supp\ w\ \Delta\ supp\ w' \qquad (7)$$

$$\forall t, w(t)\le w'(t)\Leftrightarrow supp\ w\subset supp\ w' \qquad (8)$$

2.7 **Definition** Let the function $w:\boldsymbol{R}\to\boldsymbol{B}$. The *left limit* and the *right limit*

functions of $w$, $\boldsymbol{R} \ni t \mapsto w(t-0) \in \boldsymbol{B}, \boldsymbol{R} \ni t \mapsto w(t+0) \in \boldsymbol{B}$ are given by:

$$\forall t, \exists \varepsilon > 0, \forall \xi \in (t-\varepsilon, t), w(\xi) = w(t-0) \quad (1)$$

$$\forall t, \exists \varepsilon > 0, \forall \xi \in (t, t+\varepsilon), w(\xi) = w(t+0) \quad (2)$$

We say that $w$ *has limits* or, equivalently, that the limits $w(t-0), w(t+0)$ *exist*. When $t$ is fixed, the numbers $w(t-0), w(t+0)$ are called the *left limit*, respectively the *right limit of* $w$ *in* $t$.

2.8 **Remark** $w$ defines in a unique manner the left limit and the right limit functions $w(t-0)$, $w(t+0)$ (this happens because $w$ is a function); there exist functions $w$ for which one or both of these functions do not exist.

2.9 **Notation** $$L = \{w \mid w(t-0), w(t+0) \text{ exist}\} \quad (1)$$

2.10 **Definition** The functions $\overline{w(t-0)} \cdot w(t), w(t-0) \cdot \overline{w(t)}, \overline{w(t+0)} \cdot w(t)$, $w(t+0) \cdot \overline{w(t)}$ are called the (*left*, respectively *right*) *semi-derivatives* of $w$ and the functions

$$Dw(t) = \overline{w(t-0)} \cdot w(t) \cup w(t-0) \cdot \overline{w(t)} = w(t-0) \oplus w(t) \quad (1)$$

$$D^* w(t) = \overline{w(t+0)} \cdot w(t) \cup w(t+0) \cdot \overline{w(t)} = w(t+0) \oplus w(t) \quad (2)$$

are called the (*left*, respectively *right*) *derivatives* of $w$. By fixing $t$, the previous numbers are called the *semi-derivatives* and the *derivatives* of $w$ in $t$.

2.11 **Definition** If $Dw = \mathbf{0}$ ($D^* w = \mathbf{0}$), then $w$ is called *left continuous* (*right continuous*). When fixing $t$, we get the *left continuity* (the *right continuity*) of $w$ in $t$.

2.12 **Theorem** The next conditions are equivalent for $w : \boldsymbol{R} \to \boldsymbol{B}$:

a) the unbounded family $0 \le t_0 < t_1 < t_2 < \ldots$ exists so that

$$w(t) = w(t_0) \cdot \chi_{[t_0, t_1)}(t) \oplus w(t_1) \cdot \chi_{[t_1, t_2)}(t) \oplus \ldots \quad (1)$$

b) $w$ satisfies:

b.1) $w \in L$

b.2) $supp\, w \subset [0, \infty)$

b.3) $D^* w = \mathbf{0}$ (2)

**Sketch of the proof** $a) \Rightarrow b)$ It is shown that $\forall t$, three possibilities exist:

i) $t < t_0$; then $w(t-0) = w(t+0) = w(t) = 0$

ii) $t \ge t_0$ and $\exists k, t = t_k$; then $w(t-0) = \begin{cases} w(t_{k-1}), k \ge 1 \\ 0, \quad k = 0 \end{cases}$, $w(t+0) = w(t_k)$

iii) $t \ge t_0$ and $\exists k, t \in (t_k, t_{k+1})$; then $w(t-0) = w(t+0) = w(t_k)$

$b) \Rightarrow a)$ It is shown that b.1) implies the existence of an upper and lower unbounded family $... < t_{-1} < t_0 < t_1 < ...$ satisfying the property that $\forall t$,

$$w(t) = ... \oplus w(t_{-1}) \cdot \chi_{\{t_{-1}\}}(t) \oplus w(\frac{t_{-1}+t_0}{2}) \cdot \chi_{(t_{-1},t_0)}(t) \oplus$$

$$\oplus w(t_0) \cdot \chi_{\{t_0\}}(t) \oplus w(\frac{t_0+t_1}{2}) \cdot \chi_{(t_0,t_1)}(t) \oplus w(t_1) \cdot \chi_{\{t_1\}}(t) \oplus ... \quad (3)$$

b.2), b.3) and equation (3) imply a).

2.13 **Definition** A function $w$ satisfying one of the equivalent conditions 2.12 is called *signal*.

2.14 **Remark** Our notion of signal corresponds to what is called in [5] and elsewhere *piecewise constant* signal or *non-zeno* signal or *finite variability* signal (we refer to the existence in the sketch of proof of Theorem 2.12 of the consequences that for all $t > 0$, the sets $\{k \mid t_k < t\}$ at a) and $\{k \mid -t < t_k < t\}$ at b) are finite: this is the *finite variability*; in previous works, we have called this property *local finiteness*). We avoid in this manner modelling non-inertial circuits.

On the other hand, the signals have limits, they have a non-negative support and they are right continuous. We associate these conditions with (strong) *non-anticipation*, where the output at a moment $t$ does not depend on the input (at moment $t$ and) at later moments. Dually, the functions with limits, with a non-positive support and left continuous are called *signals** and they are associated with (strong) anticipation. We shall not make use of this duality in the paper.

2.15 **Notation** $$S = \{w \mid w \textit{ is signal}\} \quad (1)$$

2.16 **Conventions** We make the following conventions when drawing graphics of $\boldsymbol{R} \to \boldsymbol{B}$ functions:

- the two values 0 and 1 are not written on the vertical axis; the low level is understood to be 0 and the high level is understood to be 1
- the vertical lines are drawn, even if they do not belong to the graphic
- we put bullets on the graphic, showing which point belongs to it

when the function switches.

2.17 **Example** We have the situation from the next figure, where $w \in S$ :

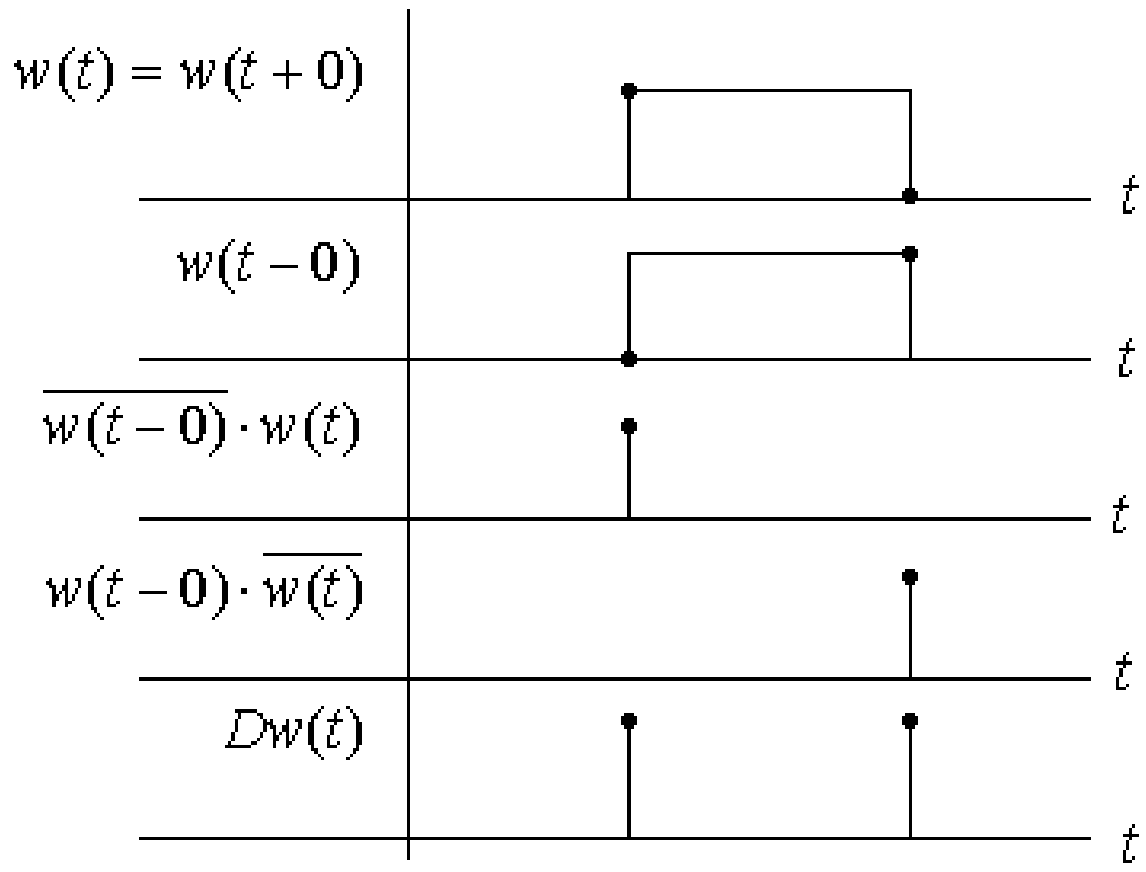

Fig 2

2.18 **Notation** We put $\tau^d : \boldsymbol{R} \to \boldsymbol{R}$ for the *translation* with $d \in \boldsymbol{R}$:

$$\tau^d(t) = t - d \qquad (1)$$

2.19 **Theorem** Let $w \in S$ an arbitrary signal and $d \in \boldsymbol{R}$. The following statements are true:

a) $w \circ \tau^d$ has limits and is right continuous

b) $w \circ \tau^d \in S \Leftrightarrow supp\ w \circ \tau^d \subset [0, \infty)$

c) if $d \geq 0$, then $w \circ \tau^d \in S$.

**Proof** a) results from 2.12 a); b) results from 2.12 b) and from a); and for c), we take in consideration b) and the fact that

$$supp\ w \circ \tau^d = \{t \mid w(t-d) = 1\} = \{t + d \mid w(t) = 1\} \subset [0, \infty) \qquad (1)$$

2.20 **Definition** The signal $w \in S$ is said to *have a limit when* $t$ *tends to* $\infty$ ($w$ is called *ultimately constant* in [6]) if the next condition holds:

$$\exists t_1, \forall t > t_1, w(t) = w(t_1) \qquad (1)$$

We also say that the limit of $w$ *when* $t$ *tends to* $\infty$ *exists.* The number $w(t_1)$ is called the *limit of* $w$ *when* $t$ *tends to* $\infty$.

2.21 **Notation** The satisfaction of the previous property is noted $\exists \lim_{t \to \infty} w(t)$ and the limit $w(t_1)$ is noted $\lim_{t \to \infty} w(t)$.

2.22 **Definition** For $w : \boldsymbol{R} \to \boldsymbol{B}$ and $A \subset \boldsymbol{R}$, we define

$$\bigcap_{\xi \in A} w(\xi) = \begin{cases} 0, \exists \xi \in A, w(\xi) = 0 \\ 1, else \end{cases} \qquad \bigcap_{\xi \in \emptyset} w(\xi) = 1 \tag{1}$$

$$\bigcup_{\xi \in A} w(\xi) = \begin{cases} 1, \exists \xi \in A, w(\xi) = 1 \\ 0, else \end{cases} \qquad \bigcup_{\xi \in \emptyset} w(\xi) = 0 \tag{2}$$

2.23 **Lemma** Let $w \in S$ and the numbers $0 \le m \le d$. The functions

$$\Phi(t) = \bigcap_{\xi \in [t-d, t-d+m]} w(\xi) \tag{1}$$

$$\Psi(t) = \bigcup_{\xi \in [t-d, t-d+m]} w(\xi) \tag{2}$$

are signals and they satisfy

$$\Phi(t-0) = w(t-d-0) \cdot \bigcap_{\xi \in [t-d, t-d+m)} w(\xi) \tag{3}$$

$$\Psi(t-0) = w(t-d-0) \cup \bigcup_{\xi \in [t-d, t-d+m)} w(\xi) \tag{4}$$

**Proof** If $m = 0$ then $\Phi(t) = \Psi(t) = w(t-d)$ and we use 2.19 c) and 2.22.

We suppose now that $m > 0$. We refer to $\Phi(t)$ and to the definition 2.12 b), for which the property 2.12 b.2), i.e. $supp\, \Phi \subset supp\, w \circ \tau^d \subset [0, \infty)$ is obvious. Let $t$ arbitrary and fixed. The left limit of $w$ in $t-d$ shows the existence of $\varepsilon_1 > 0$ with

$$\forall \xi \in [t-d-\varepsilon_1, t-d), w(\xi) = w(t-d-0) \tag{5}$$

and the left limit of $w$ in $t-d+m$ shows the existence of $\varepsilon_2 > 0$ so that

$$\forall \xi \in [t-d+m-\varepsilon_2, t-d+m), w(\xi) = w(t-d+m-0) \tag{6}$$

For any $0 < \varepsilon < \min(\varepsilon_1, \varepsilon_2, m)$ we infer

$$\Phi(t-\varepsilon) = \bigcap_{\xi \in [t-d-\varepsilon, t-d+m-\varepsilon]} w(\xi) = \bigcap_{\xi \in [t-d-\varepsilon, t-d)} w(\xi) \cdot \bigcap_{\xi \in [t-d, t-d+m-\varepsilon]} w(\xi) =$$

$$= w(t-d-0) \cdot \bigcap_{\xi \in [t-d, t-d+m-\varepsilon]} w(\xi) =$$

$$= w(t-d-0)\cdot \bigcap_{\xi\in[t-d,t-d+m-\varepsilon]} w(\xi) \cdot w(t-d+m-0) =$$

$$= w(t-d-0)\cdot \bigcap_{\xi\in[t-d,t-d+m-\varepsilon]} w(\xi) \cdot \bigcap_{\xi\in[t-d+m-\varepsilon,t-d+m)} w(\xi) =$$

$$= w(t-d-0)\cdot \bigcap_{\xi\in[t-d,t-d+m)} w(\xi) \qquad (7)$$

Because the value of $\Phi(t-\varepsilon)$ does not depend on $\varepsilon$, we get $\Phi(t-\varepsilon)=\Phi(t-0)$ and because $t$ was arbitrary, (3) is proved.

The right continuity of $w$ in $t-d$ shows the existence of $\varepsilon_3 > 0$ so that

$$\forall\xi\in[t-d,t-d+\varepsilon_3], w(\xi) = w(t-d) \qquad (8)$$

and on the other hand the right continuity of $w$ in $t-d+m$ shows the existence of $\varepsilon_4 > 0$ with

$$\forall\xi\in[t-d+m,t-d+m+\varepsilon_4], w(\xi) = w(t-d+m) \qquad (9)$$

We take some $0<\varepsilon'<\min(\varepsilon_3,\varepsilon_4,m)$ for which we have

$$\Phi(t+\varepsilon') = \bigcap_{\xi\in[t-d+\varepsilon',t-d+m+\varepsilon']} w(\xi) = \bigcap_{\xi\in[t-d+\varepsilon',t-d+m]} w(\xi) \cdot \bigcap_{\xi\in[t-d+m,t-d+m+\varepsilon']} w(\xi) =$$

$$= \bigcap_{\xi\in[t-d+\varepsilon',t-d+m]} w(\xi) \cdot w(t-d+m) = \bigcap_{\xi\in[t-d+\varepsilon',t-d+m]} w(\xi) =$$

$$= w(t-d)\cdot \bigcap_{\xi\in[t-d+\varepsilon',t-d+m]} w(\xi) =$$

$$= \bigcap_{\xi\in[t-d,t-d+\varepsilon']} w(\xi) \cdot \bigcap_{\xi\in[t-d+\varepsilon',t-d+m]} w(\xi) = \bigcap_{\xi\in[t-d,t-d+m]} w(\xi) = \Phi(t) \qquad (10)$$

The fact that $\Phi(t+\varepsilon')$ does not depend on $\varepsilon'$ shows that $\Phi(t+\varepsilon') = \Phi(t+0) = \Phi(t)$ and because $t$ was arbitrary, 2.12 b.3) is true. Moreover, 2.12 b.1) is true. $\Phi$ is signal.

Proving the same properties for $\Psi$ is made analoguely by duality.

**3. Stability. Delays**

3.1 **Definition** Let $u, x \in S$ . The property

$$\exists \lim_{t\to\infty} u(t) \Rightarrow \exists \lim_{t\to\infty} x(t) \; and \; (\lim_{t\to\infty} u(t) = \lim_{t\to\infty} x(t))$$

is called the *stability condition* (SC); $u$ is called *input* (or *control*) and $x$ is called *state* (or *output*). We say that the couple $(u,x)$ satisfies SC.

3.2 **Definition** A *delay condition* (DC) is a function that associates to each input $u \in S$ a non-empty set of states $Sol(u) \subset S$ so that $\forall x \in Sol(u)$, $(u, x)$ satisfies SC.

3.3 **Remark** SC states that if a time instant $t_1$ from which the input becomes constant exists, then a time instant $t_2$ from which the state becomes constant exists and moreover $x(t_2) = u(t_1)$ and DC restricts for each $u$ the set $\{x \mid (u, x)\ satisfies\ SC\}$ to $Sol(u)$, generally by a system of equations or inequations with the solutions $x \in Sol(u)$. Thus SC is related with the computation of the identity function $\boldsymbol{B} \ni \lim_{t \to \infty} u(t) \mapsto \lim_{t \to \infty} x(t) \in \boldsymbol{B}$ that is made by accident, anticipatory, if $t_2 < t_1$ and on purpose, non-anticipatory, if $t_2 \geq t_1$. In the first case, the *transmission delay for transitions* or shortly the *delay* in the computation of the identity on $\boldsymbol{B}$ is considered to be 0 and in the last case, the delay is $t_2 - t_1 \geq 0$.

3.4 **Example** In Fig 3:

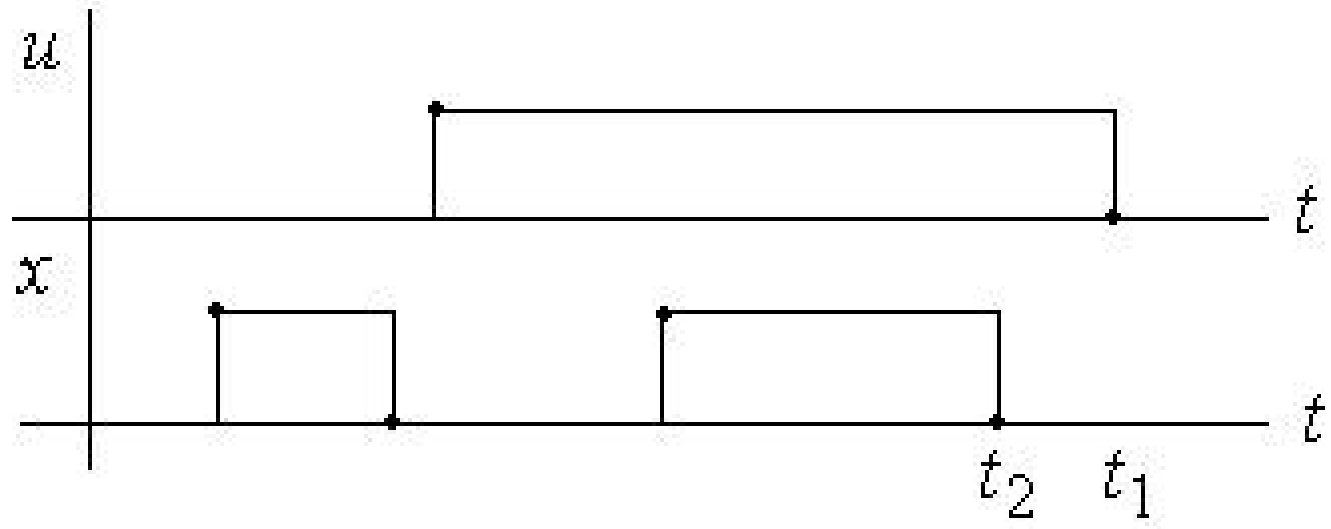

Fig 3

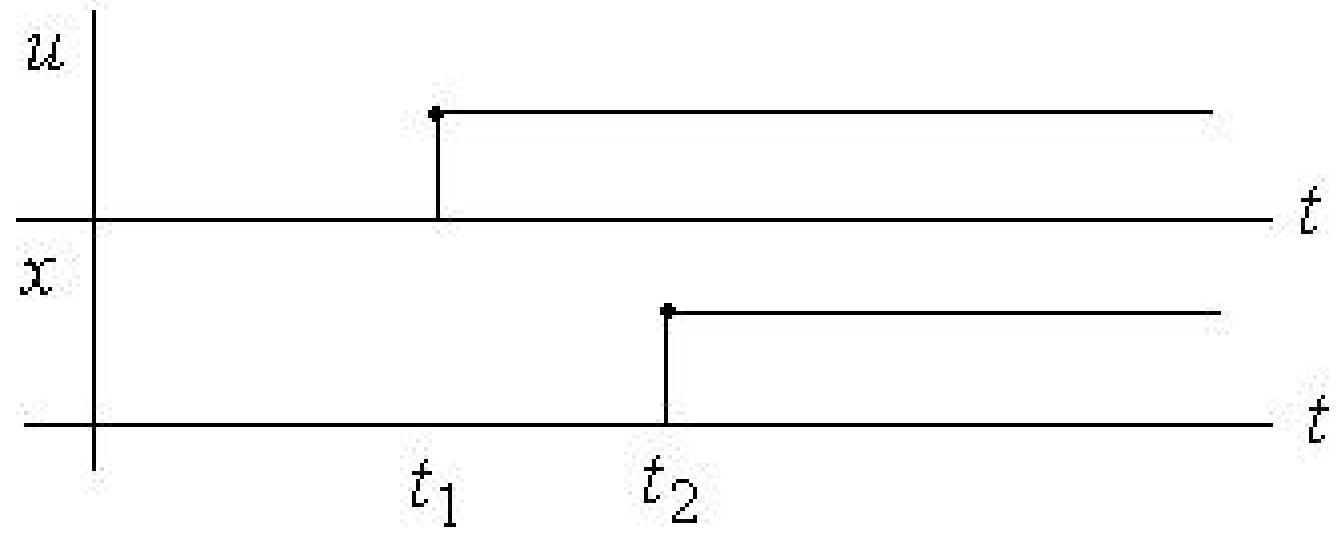

Fig 4

we have $t_2 < t_1$ and $\lim_{t\to\infty} u(t) = \lim_{t\to\infty} x(t) = 0$, while the case from Fig 4 is the one of a delay $t_2 - t_1 \geq 0$ in the computation of the identity function and $\lim_{t\to\infty} u(t) = \lim_{t\to\infty} x(t) = 1$.

3.5 **Remark** The input-output association $u \mapsto x$ is not of (uni-valued) function type, in general, because of '*delay variations due to statistical fluctuations in the fabrication process, variations in ambient temperature, power supply etc*' [3].

3.6 **Notation** We note with

$$Sol_{SC}(u) = \{x \mid (u,x) \text{ satisfies } SC\} \qquad (1)$$

the set of these $x$ that satisfy SC.

3.7 **Remark** If $\lim_{t\to\infty} u(t)$ does not exist, then $Sol_{SC}(u) = S$.

3.8 **Definition** A *delay element* (DE) (or *delay circuit* or *delay buffer*) is a subset $D \subset S \times S$ fulfilling the following conditions:

i) $\forall u, \exists x, (u,x) \in D$

ii) $\forall (u,x) \in D$, we have $x \in Sol_{SC}(u)$

iii) $\forall u, \forall x, \forall d \in \boldsymbol{R}$,

$$(u \circ \tau^d \in S \text{ and } (u,x) \in D) \Rightarrow (x \circ \tau^d \in S \text{ and } (u \circ \tau^d, x \circ \tau^d) \in D)$$

3.9 **Definition** The identity function

$$I = \{(u,u) \mid u \in S\} \qquad (1)$$

is called the *trivial* DE, or the *wire*.

3.10 **Definition** If in Definition 3.8 we replace i) with the more restrictive request

i') $\forall u, \exists! x, (u,x) \in D$

- $\exists! x$ is the notation for '*a unique x exists*' – then the DE $D$ is called *deterministic*. A DE that is not deterministic is called *non-deterministic*.

3.11 **Remark** 3.8 i) means that a DE is a total relation and the sense of 3.8 ii) was discussed at 3.3. In fact, 3.8 i), ii) state that the function $u \mapsto \{x \mid (u,x) \in D\}$ is a DC. 3.8 iii) states that $D$ is *time invariant*[3]. Let us observe that the identity function $I = 1_S$ satisfies the conditions 3.8, so that Definition 3.9 has sense. On the other hand, Definition 3.10 contains that special case when $D$ is a functional (i.e. uni-valued) relation and the sets $\{x \mid (u,x) \in D\}$ have one element (generally the solution of a system is unique) for all $u$.

3.12 **Notations** In electrical engineering, the non-trivial DE is symbolized under one of the forms from Fig 5

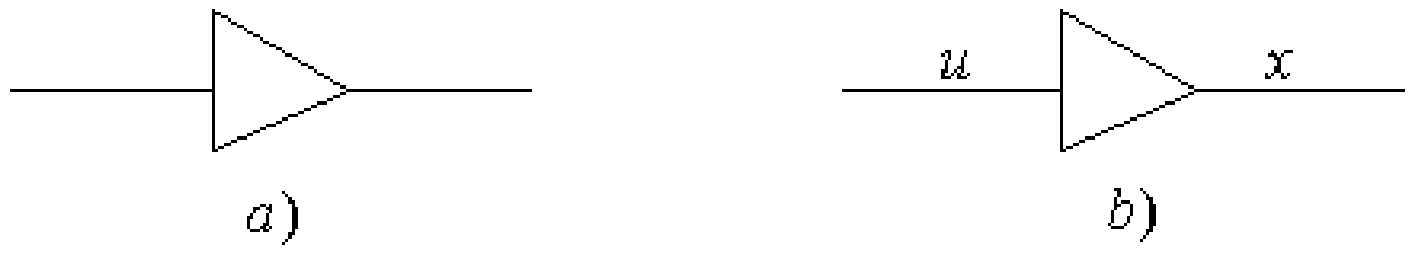

Fig 5

and the wire is noted like in Fig 6

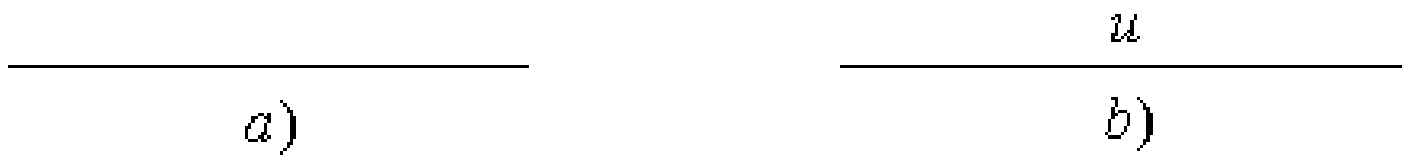

Fig 6

3.13 **Lemma** Let $D$ be an arbitrary DE. We have the next equivalence:

$$\forall u,x \in S, \forall d \geq 0, (u,x) \in D \Leftrightarrow (u \circ \tau^d, x \circ \tau^d) \in D \qquad (1)$$

| | | | |
|---|---|---|---|
| **Proof** | $u,x \in S$ *and* $d \geq 0$ | (hypothesis) | (2) |
| $\Rightarrow$ | $u \circ \tau^d \in S$ | ((2)+2.19 c)) | (3) |
| | $(u,x) \in D$ | (hypothesis $\Rightarrow$) | (4) |
| | $x \circ \tau^d \in S$ *and* $(u \circ \tau^d, x \circ \tau^d) \in D$ | ((3)+(4)+3.8 iii)) | (5) |
| $\Leftarrow$ | $(u \circ \tau^d) \circ \tau^{-d} \in S$ | (2) | (6) |

[3] In R.E. Kalman, P. L. Falb, M. A. Arbib, Topics in Mathematical System Theory, McGraw-Hill Inc. 1969, the notion of *constant*, or *time invariant system* is used. The DE's are systems in the sense of the system theory.

$$(u\circ\tau^{d},x\circ\tau^{d})\in D \qquad \text{(hypothesis } \Leftarrow) \qquad (7)$$

$$(x\circ\tau^{d})\circ\tau^{-d}\in S \; and \; (u,x)\in D \qquad ((6)+(7)+3.8 \text{ iii})) \qquad (8)$$

3.14 **Notation** For the DE's $D, D'$, we note with $D''=D'\circ D$ the set

$$D''=\{(u,y)\,|\,\exists x,(u,x)\in D \; and \; (x,y)\in D'\} \qquad (1)$$

representing the composition of the relations $D', D$ (in this order).

3.15 **Definition** $D''$ is called the *serial connection* of the DE's $D'$ and $D$.

3.16 **Theorem** $D''$ is a DE and if $D, D'$ are deterministic, then $D''$ is deterministic.

**Proof** The requests 3.8 are verified step by step. We prove 3.8 iii) and we must show that for any $u, y\in S, d\in \boldsymbol{R}$, we have the implication:

$$(u\circ\tau^{d}\in S \; and \; (u,y)\in D'')\Rightarrow(y\circ\tau^{d}\in S \; and \; (u\circ\tau^{d},y\circ\tau^{d})\in D'')$$

Let $x\in S$, whose existence is guaranteed by the definition of $D''$, so that $(u,x)\in D$ and $(x,y)\in D'$. By applying 3.8 iii) to $D$, because $u\circ\tau^{d}\in S$, we have that $x\circ\tau^{d}\in S$ and $(u\circ\tau^{d},x\circ\tau^{d})\in D$. We apply again 3.8 iii) to $D'$ and we get $y\circ\tau^{d}\in S$ and $(x\circ\tau^{d},y\circ\tau^{d})\in D'$. Thus $(u\circ\tau^{d},y\circ\tau^{d})\in D''$ and this completes the proof of 3.8 iii).

The assertion concerning the determinism is equivalent with saying that the composition of two functions is a function.

3.17 **Corollary** For a DE $D$

$$D\circ I=I\circ D=D \qquad (1)$$

is true, i.e. the wire is the neuter element relative to the serial connection.

## 4. Bounded Delays

4.1 **Definition** Let the signals $u, x$ and the numbers $0\le m_r\le d_r$, $0\le m_f\le d_f$. The system:

$$\bigcap_{\xi\in[t-d_r,t-d_r+m_r]} u(\xi) \le x(t)\le \bigcup_{\xi\in[t-d_f,t-d_f+m_f]} u(\xi) \qquad (1)$$

is called the *bounded delay condition* (BDC); $u, x$ are the *input* (or the *control*), respectively the *state* (or *output*); $m_r, m_f$ are the (*rising*, *falling*) *memories* (or the *thresholds for cancellation* or the *inertia parameters*) and $d_r, d_f$ are the (*rising, falling*) *upper bounds of the* (*transmission*) *delays* (*for transitions*). The differences $d_f - m_f$, respectively $d_r - m_r$ are called the (*rising, falling*) *lower bounds of the* (*transmission*) *delays* (*for transitions*). We say that the tuple $(u, m_r, d_r, m_f, d_f, x)$ satisfies BDC. A DC that is not bounded is called *unbounded*.

4.2 **Notation** We put

$$Sol_{BDC}(u, m_r, d_r, m_f, d_f) = \{x \mid (u, m_r, d_r, m_f, d_f, x) \text{ satisfies BDC}\} \quad (1)$$

for the set of the solutions of BDC.

4.3 **Theorem** Let $0 \le m_r \le d_r, 0 \le m_f \le d_f$ be given. The following statements are equivalent:

a) $$\forall u, \bigcap_{\xi \in [t-d_r, t-d_r+m_r]} u(\xi) \le \bigcup_{\xi \in [t-d_f, t-d_f+m_f]} u(\xi) \quad (1)$$

b) $$\forall u, \bigcap_{\xi \in [t-d_r, t-d_r+m_r]} u(\xi) \cdot \bigcap_{\xi \in [t-d_f, t-d_f+m_f]} \overline{u(\xi)} = 0 \quad (2)$$

c) $$d_r - m_r \le d_f \text{ and } d_f - m_f \le d_r \quad (3)$$

d) $$\forall u, Sol_{BDC}(u, m_r, d_r, m_f, d_f) \ne \emptyset \quad (4)$$

**Proof** $a) \Leftrightarrow b)$: $$\forall u, \bigcap_{\xi \in [t-d_r, t-d_r+m_r]} u(\xi) \le \bigcup_{\xi \in [t-d_f, t-d_f+m_f]} u(\xi)$$

$$\Leftrightarrow \forall u, \overline{\bigcap_{\xi \in [t-d_r, t-d_r+m_r]} u(\xi)} \cup \overline{\bigcap_{\xi \in [t-d_f, t-d_f+m_f]} \overline{u(\xi)}} = 1$$

$$\Leftrightarrow \forall u, \overline{\bigcap_{\xi \in [t-d_r, t-d_r+m_r]} u(\xi) \cdot \bigcap_{\xi \in [t-d_f, t-d_f+m_f]} \overline{u(\xi)}} = 1$$

$$\Leftrightarrow \forall u, \bigcap_{\xi \in [t-d_r, t-d_r+m_r]} u(\xi) \cdot \bigcap_{\xi \in [t-d_f, t-d_f+m_f]} \overline{u(\xi)} = 0 \quad (5)$$

$b) \Leftrightarrow c)$: $$\forall u, \bigcap_{\xi \in [t-d_r, t-d_r+m_r]} u(\xi) \cdot \bigcap_{\xi \in [t-d_f, t-d_f+m_f]} \overline{u(\xi)} = 0$$

$$\Leftrightarrow \forall t, [t - d_r, t - d_r + m_r] \wedge [t - d_f, t - d_f + m_f] \ne \emptyset$$

(because if $\exists t, [t - d_r, t - d_r + m_r] \wedge [t - d_f, t - d_f + m_f] = \emptyset$, then $u$ exists so that $\forall \xi \in [t - d_r, t - d_r + m_r], u(\xi) = 1$ and $\forall \xi \in [t - d_f, t - d_f + m_f]$, $u(\xi)=0$; conversely, if $\forall t,[t-d_r,t-d_r+m_r]\wedge[t-d_f,t-d_f+m_f]\neq\emptyset$, then $\forall u,\forall t,\forall\xi\in[t-d_r,t-d_r+m_r]\wedge[t-d_f,t-d_f+m_f]$, $u(\xi)=1$ and $u(\xi)=0$ cannot be both true)

$$\Leftrightarrow\forall t,\neg(t-d_r+m_r<t-d_f\ or\ t-d_f+m_f<t-d_r)$$

$$\Leftrightarrow\forall t,t-d_r+m_r\geq t-d_f\ and\ t-d_f+m_f\geq t-d_r$$

$$\Leftrightarrow d_r-m_r\leq d_f\ and\ d_f-m_f\leq d_r \qquad (6)$$

$a)\Leftrightarrow d)$ is obvious if we take into account Lemma 2.23.

4.4 **Definition** The properties 4.3 are called the *consistency condition* (CC).

4.5 **Remark** For any $0\leq m_r\leq d_r$, $0\leq m_f\leq d_f$ we have $Sol_{BDC}(\mathbf{0},m_r,d_r,m_f,d_f)=\{\mathbf{0}\}$ and if $u\neq\mathbf{0}$ and CC is fulfilled, then $Sol_{BDC}(u,m_r,d_r,m_f,d_f)$ is either a one element set or an infinite set.

4.6 **Remark** The situation expressed by BDC is shown in the following figure where we have supposed that $m_r<\tau$ and that CC is fulfilled: $d_r-m_r\leq d_f$, $d_f-m_f\leq d_r$. We observe that if $u$ is 1 for more than $m_r$ time units then $x$ becomes 1 with a delay of $d\in[d_f-m_f,d_r]$ time units and, in a dual manner, that if $u$ is 0 for more than $m_f$ time units then $x$ becomes 0 with a delay of $d \in [d_r - m_r, d_f]$ time units. See also the informal definition of the bounded delays from [2], [3] in our Introduction at the classification of the delays, item **II** b).

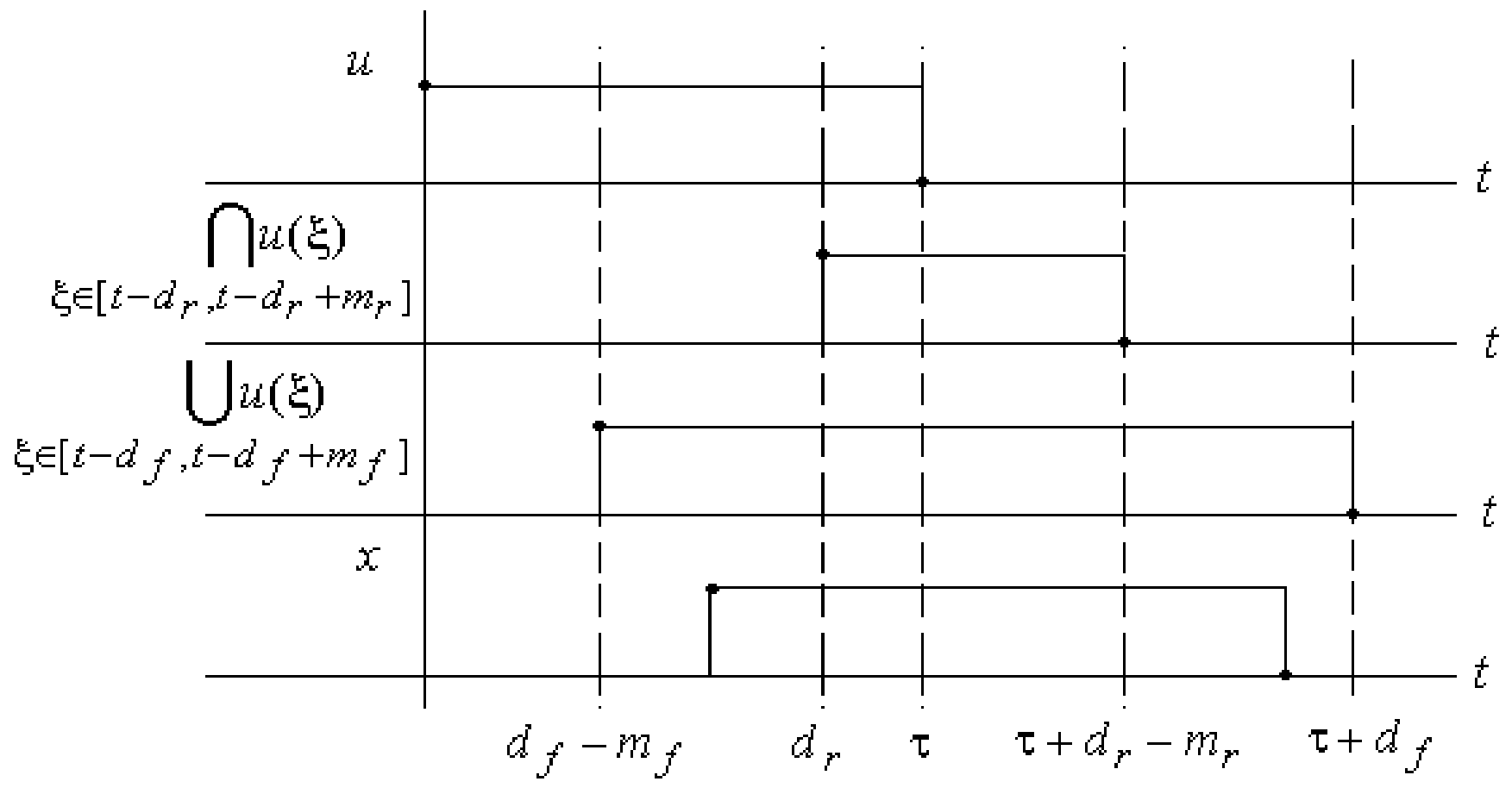

Fig 7

4.7 **Remark** Two interpretations of BDC exist, the positive and the negative interpretation.

The *positive interpretation* of BDC is the following: it is natural that the value of the output be arbitrary when the input is not sufficiently persistent (in Fig 7 from 4.6, for $t \in [d_f - m_f, d_r) \vee [\tau + d_r - m_r, \tau + d_f))$. This fact could also be modelled by the replacement of the set $\{0,1\}$ with $\{0, \frac{1}{2}, 1\}$ -some authors do so- but this has algebraical disadvantages, because the set $\{0, \frac{1}{2}, 1\}$ is poorer algebraically than $\{0,1\}$. Non-determinism is another way (ours!) of solving this problem.

The *negative interpretation* of BDC: it is not natural that at some time moment, $t_1$ for example, we may have $x(t_1 - 0) \cdot \overline{x(t_1)} = 1$, while $\forall t \leq t_1, u(t-0) \cdot \overline{u(t)} = 0$ in the sense that $x$ is allowed to switch in a manner that is not compatible with the (more or less) (left) local behaviour of $u$.

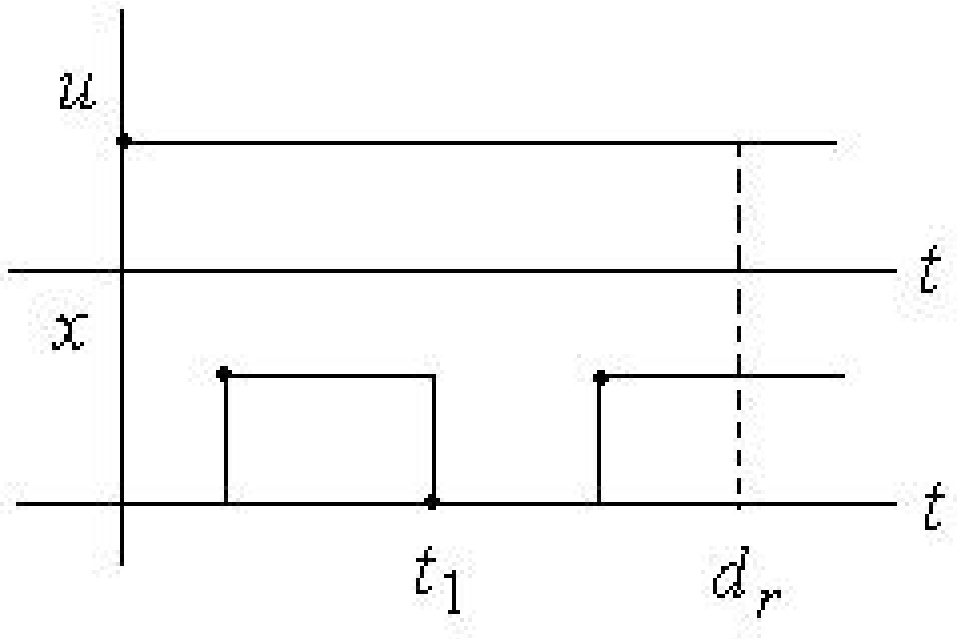

Fig 8

In other words, pulses on the output may exist that do not reproduce the pulses on the input.

On the other hand, we observe that the pulses that are shorter than or equal to $m_r$, respectively shorter than or equal to $m_f$ on the input are not necessarily transmitted to the output, thus BDC implies inertia when $m_r > 0$ or $m_f > 0$; we have called this type of inertia '*trivalent*', because it may be interpreted in the trivalent logic and '*relative*' because it relates $x$ with $u$.

4.8 **Remark** Let us suppose that in BDC the upper and the lower bounds of the delays coincide:

$$d_r = d_f - m_f \qquad (1)$$

$$d_f = d_r - m_r \qquad (2)$$

representing the situation when, see Fig 7 from 4.6, in 4.1 (1) and in 4.3 (1), (3) the inequalities are replaced by equalities.

We add (1) and (2) and we get $m_r + m_f = 0$, i.e.

$$m_r = m_f = 0 \qquad (3)$$

$$d_r = d_f = d \qquad (4)$$

and BDC takes the form

$$x(t) = u(t-d) \qquad (5)$$

We shall treat this special case called the fixed delay condition in the next chapter. We shall also meet equation (5) as a special case of bounded inertial delay condition, more specifically of deterministic inertial delay condition (chapter 8) where there is no inertia (equation (3)) and the transmission delays for transitions are equal (equation (4)).

4.9 **Theorem** For any $u$ and any $0 \le m_r \le d_r, 0 \le m_f \le d_f$, we have:

$$CC \Rightarrow (Sol_{BDC}(u, m_r, d_r, m_f, d_f) \subset Sol_{SC}(u)) \qquad (1)$$

**Proof** If CC is fulfilled and if $\exists \lim\limits_{t\to\infty} u(t)$, then $x \in Sol_{BDC}(u, m_r, d_r, m_f, d_f)$ and $t_1 \ge 0$ exist so that

$$\forall t \ge t_1, \quad \bigcap_{\xi\in[t-d_r, t-d_r+m_r]} u(\xi) = x(t) = \bigcup_{\xi\in[t-d_f, t-d_f+m_f]} u(\xi) = \lim_{t\to\infty} u(t) \qquad (2)$$

4.10 **Theorem** (*of representation of the solutions of BDC*). Let $u \in S$ and the numbers $0 \le m_r \le d_r$, $0 \le m_f \le d_f$ so that CC is fulfilled. The following statements are equivalent:

a) $x \in Sol_{BDC}(u, m_r, d_r, m_f, d_f)$

b) $\exists y \in S$ so that $x$ is given by

$$x(t) = \bigcap_{\xi\in[t-d_r, t-d_r+m_r]} u(\xi) \cup y(t) \cdot \bigcup_{\xi\in[t-d_f, t-d_f+m_f]} u(\xi) \qquad (1)$$

**Proof** The satisfaction of the consistency condition makes that, at a certain time instant $t$, we have one of the cases i), ii), iii) true:

i) $$\bigcap_{\xi\in[t-d_r,t-d_r+m_r]} u(\xi) = \bigcup_{\xi\in[t-d_f,t-d_f+m_f]} u(\xi) = 0 \qquad (2)$$

ii) $$\bigcap_{\xi\in[t-d_r,t-d_r+m_r]} u(\xi) = 0, \quad \bigcup_{\xi\in[t-d_f,t-d_f+m_f]} u(\xi) = 1 \qquad (3)$$

iii) $$\bigcap_{\xi\in[t-d_r,t-d_r+m_r]} u(\xi) = \bigcup_{\xi\in[t-d_f,t-d_f+m_f]} u(\xi) = 1 \qquad (4)$$

$a) \Rightarrow b)$ Let $x \in Sol_{BDC}(u, m_r, d_r, m_f, d_f)$. In the cases i), iii) $y(t)$ has arbitrary values and in case ii) $y(t) = x(t)$.

$b) \Rightarrow a)$ In all the cases i), ii), iii) BDC is obviously fulfilled for any $y$.

4.11 **Theorem** If $0 \le m_r \le d_r$ and $0 \le m_f \le d_f$ satisfy CC, then the set

$$D = \{(u,x) \mid u \in S, x \in Sol_{BDC}(u, m_r, d_r, m_f, d_f)\} \qquad (1)$$

is a DE.

**Proof** 3.8 i) is true because CC is fulfilled and 3.8 ii) was proved in Theorem 4.9. We check 3.8 iii).

We prove that

$$(u \circ \tau^d \in S \text{ and } (u,x) \in D) \Rightarrow x \circ \tau^d \in S$$

and let $d \in \mathbf{R}$ fixed. The statement $u \circ \tau^d \in S$, equivalent with $u(t-d) = 0, t < 0$ gives

$$u(t) = 0, t < -d \Leftrightarrow u(t) \le \chi_{[-d,\infty)}(t) \qquad (2)$$

$$\bigcup_{\xi\in[t-d_f,t-d_f+m_f]} u(\xi) \le$$

$$\le \bigcup_{\xi\in[t-d_f,t-d_f+m_f]} \chi_{[-d,\infty)}(\xi) = \chi_{[d_f-m_f-d,\infty)}(t) \le \chi_{[-d,\infty)}(t) \qquad (3)$$

Because $x \in Sol_{BDC}(u, m_r, d_r, m_f, d_f)$ we have

$$x(t) \le \bigcup_{\xi\in[t-d_f,t-d_f+m_f]} u(\xi) \le \chi_{[-d,\infty)}(t) \qquad (4)$$

i.e. $x(t) = 0, t < -d$ thus $x \circ \tau^d \in S$.

We prove now that

$$(u\circ\tau^d \in S \text{ and } (u,x)\in D)\Rightarrow (u\circ\tau^d, x\circ\tau^d)\in D$$

$$\bigcap_{\xi\in[t-d_r,t-d_r+m_r]}(u\circ\tau^d)(\xi)=\bigcap_{\xi\in[t-d_r,t-d_r+m_r]}u(\xi-d)=$$

$$=\bigcap_{\xi+d\in[t-d_r,t-d_r+m_r]}u(\xi)=\bigcap_{\xi\in[t-d-d_r,t-d-d_r+m_r]}u(\xi)\leq x(t-d)=$$

$$=(x\circ\tau^d)(t)\leq\bigcup_{\xi\in[t-d-d_f,t-d-d_f-m_f]}u(\xi)=\bigcup_{\xi+d\in[t-d_f,t-d_f-m_f]}u(\xi)=$$

$$=\bigcup_{\xi\in[t-d_f,t-d_f-m_f]}u(\xi-d)=\bigcup_{\xi\in[t-d_f,t-d_f-m_f]}(u\circ\tau^d)(\xi)\qquad(5)$$

4.12 **Definition** Any DE $D'\subset D$, where the set $D$ was defined at 4.11 (1) is called *bounded delay element* (BDE) (*circuit*, *buffer*). The BDE $D$ itself is called *full*. A DE that is not bounded is called *unbounded*.

4.13 **Theorem** If $D,D'$ are BDE's, then their serial connection $D''=D'\circ D$ is BDE with

$$d_r''=d_r+d_r',\ d_f''=d_f+d_f'\qquad(1)$$

$$m_r''=m_r+m_r',\ m_f''=m_f+m_f'\qquad(2)$$

**Proof** Let us observe at first that the satisfaction of the consistency condition

$$d_r-m_r\leq d_f,\ d_f-m_f\leq d_r\qquad(3)$$

$$d_r'-m_r'\leq d_f',\ d_f'-m_f'\leq d_r'\qquad(4)$$

implies

$$d_r''-m_r''\leq d_f'',\ d_f''-m_f''\leq d_r''\qquad(5)$$

where $d_r'',d_f'',m_r'',m_f''$ are given by (1), (2).

Furthermore, $u$ being the input of $D$ and $D''$, $x$ being the state of $D$ and the input of $D'$ and respectively $y$ the state of $D'$ and $D''$, from

$$\bigcap_{\xi\in[t-d_r,t-d_r+m_r]}u(\xi)\leq x(t)\leq\bigcup_{\xi\in[t-d_f,t-d_f+m_f]}u(\xi)\qquad(6)$$

$$\bigcap_{\xi\in[t-d_r',t-d_r'+m_r']}x(\xi)\leq y(t)\leq\bigcup_{\xi\in[t-d_f',t-d_f'+m_f']}x(\xi)\qquad(7)$$

we have for example

$$y(t) \geq \bigcap_{\xi \in [t-d_r', t-d_r'+m_r']} x(\xi) \geq \qquad (8)$$

$$\geq \bigcap_{\xi \in [t-d_r', t-d_r'+m_r']} \bigcap_{\xi' \in [\xi-d_r, \xi-d_r+m_r]} u(\xi') = \bigcap_{\xi \in [t-d_r'-d_r, t-d_r'-d_r+m_r'+m_r]} u(\xi)$$

4.14 **Remark** The inclusion defines an order in the set of the BDE's. If $D, D'$ are full BDE's, then $D \subset D'$ iff

$$\forall u, \quad \bigcap_{\xi \in [t-d_r', t-d_r'+m_r']} u(\xi) \leq \bigcap_{\xi \in [t-d_r, t-d_r+m_r]} u(\xi) \leq$$

$$\leq \bigcup_{\xi \in [t-d_f, t-d_f+m_f]} u(\xi) \leq \bigcup_{\xi \in [t-d_f', t-d_f'+m_f']} u(\xi) \qquad (1)$$

equivalently iff

$$\forall u, Sol_{BDC}(u, m_r, d_r, m_f, d_f) \subset Sol_{BDC}(u, m_r', d_r', m_f', d_f')$$

with the interpretation: $D, D'$ are asked to delay the input with *at least* $d_f - m_f \geq d_f' - m_f'$, respectively $d_r - m_r \geq d_r' - m_r'$ time units and with *at most* $d_r \leq d_r'$, respectively $d_f \leq d_f'$ time units.

4.15 **Remark** By passing to the left limit in BDC, we use Lemma 2.23 and we obtain the next inequalities:

$$u(t-d_r-0) \cdot \bigcap_{\xi \in [t-d_r, t-d_r+m_r)} u(\xi) \leq x(t-0) \leq$$

$$\leq u(t-d_f-0) \cup \bigcup_{\xi \in [t-d_f, t-d_f+m_f)} u(\xi) \qquad (1)$$

that are fulfilled by any $x \in Sol_{BDC}(u, m_r, d_r, m_f, d_f)$.

## 5. Fixed Delays. Constant Delays

5.1 **Definition** For $u, x \in S$ and $d \geq 0$, the relation (see Remark 4.8)

$$x(t) = u(t-d) \qquad (1)$$

is called the *fixed delay condition* (FDC) (or the *ideal delay condition* or the *pure delay condition*). $u, x$ have the same names as mentioned before and $d$ is called (*transmission*) *delay* (*for transitions*). We say that $(u, d, x)$ satisfies FDC.

5.2 **Remark** In FDC, CC is fulfilled for any $d \geq 0$ under the form $d \geq d$.

5.3 **Remark** If $u \in S$ and $d \geq 0$, then $x \in S$ like at 2.19 c).

5.4 **Notation** The one element set consisting in the solution $x$ of FDC is noted with

$$Sol_{FDC}(u,d) = \{x \mid (u,d,x) \text{ satisfies FDC}\} \quad (1)$$

5.5 **Theorem** a) $\forall u, \forall d \geq 0, Sol_{FDC}(u,d) \subset Sol_{SC}(u)$ (1)

b) For any $u$ and any numbers $0 \leq m_r \leq d_r, 0 \leq m_f \leq d_f$, we have:

$$d \in [d_r - m_r, d_r] \wedge [d_f - m_f, d_f] \Rightarrow$$

$$\Rightarrow (Sol_{FDC}(u,d) \subset Sol_{BDC}(u, m_r, d_r, m_f, d_f)) \quad (2)$$

**Proof** a) If $\exists t_1$ so that $\forall \xi \geq t_1, u(\xi) = u(t_1)$, then for any $d \geq 0$ we have $\forall \xi \geq t_1 + d, x(\xi) = u(\xi - d) = u(t_1)$. This property is a consequence however of Remark 5.2 and of Theorem 4.9.

b) If $d \in [d_r - m_r, d_r] \wedge [d_f - m_f, d_f]$, then the next statements are true:

$$\forall t, t - d_r \leq t - d \leq t - d_r + m_r \text{ and } t - d_f \leq t - d \leq t - d_f + m_f \quad (3)$$

$$\bigcap_{\xi \in [t-d_r, t-d_r+m_r]} u(\xi) \leq u(t-d) \leq \bigcup_{\xi \in [t-d_f, t-d_f+m_f]} u(\xi) \quad (4)$$

5.6 **Remark** At 5.5 b) the condition $[d_r - m_r, d_r] \wedge [d_f - m_f, d_f] \neq \emptyset$ occurs in the hypothesis; this condition coincides with CC and consequently with the fact that $Sol_{BDC}(u, m_r, d_r, m_f, d_f) \neq \emptyset$ for all $u$.

5.7 **Theorem** The set

$$I_d = \{(u, u \circ \tau^d) \mid u \in S\} \quad (1)$$

where $d \geq 0$ is a deterministic DE.

**Proof** 3.8 i) is obvious, 3.8 ii) was shown at 5.5 a) and 3.8 iii) means that (see Theorem 2.19) for all $u \in S$ and $d' \in \boldsymbol{R}$, the obvious implication

$$\{t \mid u(t-d') = 1\} \subset [0,\infty) \Rightarrow \{t \mid u(t-d'-d) = 1\} \subset [0,\infty)$$

is true. All these follow from Theorem 4.11 however.

Determinism means exactly that $Sol_{FDC}(u,d)$ is a one element set.

5.8 **Definition** $I_d$ previously defined is called the *fixed delay element* (FDE) (or the *ideal delay element* or the *pure delay element*).

5.9 **Theorem** For all $d, d' \geq 0$

$$I_d \circ I_{d'} = I_{d'} \circ I_d = I_{d+d'} \qquad (1)$$

i.e. by the serial connection of two FDE's having the delays $d, d'$ we get a FDE, having the delay $d + d'$. Moreover, the serial connection of the FDE's is commutative.

**Proof** Follows from the property

$$(u \circ \tau^d) \circ \tau^{d'} = u \circ \tau^{d+d'} \qquad (2)$$

5.10 **Remark** The set of the FDE's is organised by the serial connection as commutative monoid, where the unit is the wire.

5.11 **Theorem** Let $D$ an arbitrary DE and $I_d$ some FDE. We have:

$$D \circ I_d = I_d \circ D = \{(u, x \circ \tau^d) \mid (u, x) \in D\} \qquad (1)$$

and if $D$ is deterministic, then $D \circ I_d$ is deterministic.

**Proof** $D \circ I_d =$ (2)

$= \{(u, y) \mid \exists! x, (u, x) \in I_d \text{ and } (x, y) \in D\}$ (the determinism of $I_d$)

$= \{(u, y) \mid \exists! x, x = u \circ \tau^d \text{ and } (x, y) \in D\}$

$= \{(u, y) \mid (u \circ \tau^d, y) \in D\}$

$= \{(u, y) \mid (u, y \circ \tau^{-d}) \in D\}$ (Lemma 3.13)

$= \{(u, y) \mid \exists! x, (u, x) \in D \text{ and } x = y \circ \tau^{-d}\}$

$= \{(u, y) \mid \exists! x, (u, x) \in D \text{ and } (x, y) \in I_d\}$

$= I_d \circ D$ (the determinism of $I_d$)

5.12 **Theorem** If $D$ is BDE and $I_d$ is FDE, then $D' = D \circ I_d = I_d \circ D$ is BDE with $m_r' = m_r, d_r' = d_r + d, m_f' = m_f, d_f' = d_f + d$.

**Proof** $D' = D \circ I_d$ is a BDE with $m_r', d_r', m_f', d_f'$ like in the statement of the Theorem 4.13. The commutativity $D \circ I_d = I_d \circ D$ was proved at Theorem 5.11.

5.13 **Definition** The DC $u \mapsto Sol(u)$ is called *constant* if $d_r \geq 0, d_f \geq 0$

exist so that $\forall u, \forall x \in Sol(u)$ the next inequalities are fulfilled

$$\overline{x(t-0)} \cdot x(t) \leq u(t-d_r) \qquad (1)$$

$$x(t-0) \cdot \overline{x(t)} \leq \overline{u(t-d_f)} \qquad (2)$$

A DC that is not constant is called *variable.*

5.14 **Remark** The fixed delays are constant, because

$$\overline{x(t-0)} \cdot x(t) \leq x(t) = u(t-d) \qquad (1)$$

$$x(t-0) \cdot \overline{x(t)} \leq \overline{x(t)} = \overline{u(t-d)} \qquad (2)$$

5.15 **Remark** For the informal definitions of the fixed delays, respectively of the pure delays from [2] see also the classification of the delays in our Introduction, item II c), respectively III a).

## 6. Bivalent Relative Inertial Delays

6.1 **Lemma** Let $0 \leq \mu_r \leq \delta_r$ and $0 \leq \mu_f \leq \delta_f$ be arbitrary numbers. When $u, x$ run in $S$, the next statements are equivalent:

a) $$\overline{x(t-0)} \cdot x(t) \leq \bigcap_{\xi \in [t-\delta_r, t-\delta_r+\mu_r]} u(\xi) \qquad (1)$$

b) $$\overline{x(t-0)} \cdot x(t) \leq \overline{x(t-0)} \cdot \bigcap_{\xi \in [t-\delta_r, t-\delta_r+\mu_r]} u(\xi) \qquad (2)$$

and the next statements are also equivalent

a') $$x(t-0) \cdot \overline{x(t)} \leq \bigcap_{\xi \in [t-\delta_f, t-\delta_f+\mu_f]} \overline{u(\xi)} \qquad (3)$$

b') $$x(t-0) \cdot \overline{x(t)} \leq x(t-0) \cdot \bigcap_{\xi \in [t-\delta_f, t-\delta_f+\mu_f]} \overline{u(\xi)} \qquad (4)$$

**Proof** $a) \Rightarrow b)$ Follows by multiplying (1) with $\overline{x(t-0)}$.

$b) \Rightarrow a)$ We have

$$\overline{x(t-0)} \cdot x(t) \leq \overline{x(t-0)} \cdot \bigcap_{\xi \in [t-\delta_r, t-\delta_r+\mu_r]} u(\xi) \leq \bigcap_{\xi \in [t-\delta_r, t-\delta_r+\mu_r]} u(\xi) \qquad (5)$$

6.2 **Definition** We consider the signals $u, x$ and the numbers $0 \leq \mu_r \leq \delta_r$, $0 \leq \mu_f \leq \delta_f$. The property expressed by

i) $$\exists \lim_{t \to \infty} u(t) \Rightarrow \exists \lim_{t \to \infty} x(t) \text{ and } (\lim_{t \to \infty} u(t) = \lim_{t \to \infty} x(t))$$

ii) $$\overline{x(t-0)}\cdot x(t)\le \bigcap_{\xi\in[t-\delta_r,t-\delta_r+\mu_r]} u(\xi) \qquad (1)$$

$$x(t-0)\cdot \overline{x(t)}\le \bigcap_{\xi\in[t-\delta_f,t-\delta_f+\mu_f]} \overline{u(\xi)} \qquad (2)$$

is called the (*bivalent*) *relative inertial delay condition* (RIDC). $u,x$ have the same names like before and $\mu_r,\delta_r,\mu_f,\delta_f$ are called *inertia parameters*. We say that the tuple $(u,\mu_r,\delta_r,\mu_f,\delta_f,x)$ satisfies RIDC.

6.3 **Remark** The relative inertial delays are constant, because 6.2 ii) imply

$$\overline{x(t-0)}\cdot x(t)\le u(t-\delta_r) \qquad (1)$$

$$x(t-0)\cdot \overline{x(t)}\le \overline{u(t-\delta_f)} \qquad (2)$$

This example of constant delays is less trivial than the one of the FDC's, because we have generally $\delta_r\ne\delta_f$.

6.4 **Notation** Let

$$Sol_{RIDC}(u,\mu_r,\delta_r,\mu_f,\delta_f)=$$
$$=\{x\,|\,(u,\mu_r,\delta_r,\mu_f,\delta_f,x)\ satisfies\ RIDC\} \qquad (1)$$

the set of the solutions of RIDC.

6.5 **Remark** Because 6.2 i) coincides with SC, for any numbers $0\le\mu_r\le\delta_r, 0\le\mu_f\le\delta_f$ and any signal $u\in S$ we have

$$Sol_{RIDC}(u,\mu_r,\delta_r,\mu_f,\delta_f)\subset Sol_{SC}(u) \qquad (1)$$

6.6 **Remark** We interpret the dual inequations 6.1 (1), 6.1 (3) in the spirit of the informal definition from [2], see in our Introduction the classification of the delays, item **III** b), that we rewrite in the following manner: '*pulses shorter than or equal to* $m_r$ (respectively $m_f$) *are not transmitted and pulses strictly longer than* $m_r$ (respectively $m_f$) *may be transmitted*'. This interpretation results from the fact that for example $u=\chi_{[0,\tau)}$ implies whenever $0<\tau\le\mu_r$ that $\bigcap_{\xi\in[t-\delta_r,t-\delta_r+\mu_r]} u(\xi)=0$ and $x$ cannot switch (inequation 6.1 (1)) from 0 to 1. Such pulses are not transmitted to the output, bivalent inertia unlike the trivalent inertia from BDC where the pulses $u = \chi_{[0,\tau)}$ with $0 < \tau \le m_r$ sometimes were transmitted, sometimes were not.

6.7 **Example** We show by an example the meaning of

$$\delta_r - \mu_r \le \delta_f \, , \; \delta_f - \mu_f \le \delta_r \qquad (1)$$

analogue with 4.3 c) whose satisfaction was not asked so far, in RIDC. Let $\delta > 0$ some number and $u = \chi_{[0,\mu_r+\delta)}$ giving

$$\bigcap_{\xi \in [t-\delta_r, t-\delta_r+\mu_r]} u(\xi) = \chi_{[\delta_r,\delta_r+\delta)}(t) \qquad (2)$$

$$\bigcap_{\xi \in [t-\delta_f, t-\delta_f+\mu_f]} \overline{u(\xi)} = \chi_{(-\infty,\delta_f-\mu_f)}(t) \oplus \chi_{[\mu_r+\delta_f+\delta,\infty)}(t) \qquad (3)$$

<u>Case 1</u> (1) is true, thus $\delta_f - \mu_f \le \delta_r < \delta_r + \delta \le \mu_r + \delta_f + \delta$ and RIDC has solutions of two forms:

a) $x(t) = 0$ (4)

b) $x(t) = \chi_{[d,d')}(t)$ (5)

where $d \in [\delta_r, \delta_r + \delta)$ and $d' \in [\mu_r + \delta_f + \delta, \infty)$.

<u>Case 2</u> (1) is not satisfied and, in order to make a choice, we suppose that $[\delta_r, \delta_r + \delta) \subset [\mu_r + \delta_f + \delta, \infty)$. Then the solutions of RIDC are of the forms

a) $x(t) = 0$ (6)

b) $x(t) = \chi_{[d_1,d_2)}(t) \oplus \chi_{[d_3,d_4)}(t) \oplus ... \oplus \chi_{[d_{2n-1},d_{2n})}(t)$ (7)

where $n \ge 1$ and

$$\delta_r \le d_1 < d_2 < ... < d_{2n-1} < \delta_r + \delta \le d_{2n} \qquad (8)$$

From this point of view, we conclude that our choice for the satisfaction or not of the inequalities (1) is that -case 1 - a pulse on the input be transmitted to the output under the form of at most one pulse or -case 2 - a pulse on the input be transmitted to the output under the form of at most a finite number of pulses, having the switching moments situated in the range indicated at (8)

This undesired situation is similar with what we have called the negative interpretation of BDC at Remark 4.7.

6.8 **Remark** We avoid the previous property (of *density* or of *zenoness*), written under the general form:

$\forall \varepsilon > 0, \exists u, \exists x \in Sol_{RIDC}(u, \mu_r, \delta_r, \mu_f, \delta_f), \exists d, d' > 0,$

$$\overline{x(d-0)}\cdot x(d)=1 \; and \; x(d'-0)\cdot\overline{x(d')}=1 \; and \; |d-d'|<\varepsilon \qquad (1)$$

by asking that in RIDC the request $\delta_r-\mu_r\le\delta_f,\delta_f-\mu_f\le\delta_r$ be fulfilled. This condition has the same form like CC, but the content is different.

6.9 **Definition** In RIDC 6.2 the property $\delta_r-\mu_r\le\delta_f,\delta_f-\mu_f\le\delta_r$ is called the *non-zenoness condition* (NZC). NZC is by definition *trivial* if $\delta_r-\mu_r=\delta_f,\delta_f-\mu_f=\delta_r$ and *non-trivial* otherwise.

6.10 **Theorem** We consider $u\in S$ and $x\in Sol_{RIDC}(u,\mu_r,\delta_r,\mu_f,\delta_f)$ so that NZC be true. Then for all $d,d'\ge 0$, the next implications hold:

$$\overline{x(d-0)}\cdot x(d)=1 \; and \; x(d'-0)\cdot\overline{x(d')}=1 \; and \; d<d'\Rightarrow$$
$$\Rightarrow d'-d>\delta_f-\delta_r+\mu_r \qquad (1)$$
$$x(d-0)\cdot\overline{x(d)}=1 \; and \; \overline{x(d'-0)}\cdot x(d')=1 \; and \; d<d'\Rightarrow$$
$$\Rightarrow d'-d>\delta_r-\delta_f+\mu_f \qquad (2)$$

**Proof** We show (1). The hypothesis states

$$\bigcap_{\xi\in[d-\delta_r,d-\delta_r+\mu_r]} u(\xi) \quad = \quad \bigcap_{\xi\in[d'-\delta_f,d'-\delta_f+\mu_f]} \overline{u(\xi)} \quad =1 \qquad (3)$$

meaning that

$$[d-\delta_r,d-\delta_r+\mu_r]\wedge[d'-\delta_f,d'-\delta_f+\mu_f]=\emptyset\Leftrightarrow$$
$$\Leftrightarrow d-\delta_r+\mu_r<d'-\delta_f \; or \; d'-\delta_f+\mu_f<d-\delta_r\Leftrightarrow$$
$$\Leftrightarrow d'-d>\delta_f-\delta_r+\mu_r \; or \; d'-d<\delta_f-\mu_f-\delta_r\Leftrightarrow$$
$$\Leftrightarrow d'-d>\delta_f-\delta_r+\mu_r \qquad (4)$$

6.11 **Remark** Let $0\le\mu_r\le\delta_r$, $0\le\mu_f\le\delta_f$ arbitrary. We have $Sol_{RIDC}(\mathbf{0},\mu_r,\delta_r,\mu_f,\delta_f)=\{\mathbf{0}\}$ and if $u\ne\mathbf{0}$ and NZC is true, then $Sol_{RIDC}(u,\mu_r,\delta_r,\mu_f,\delta_f)=\{\mathbf{0}\}$ and $Sol_{RIDC}(u,\mu_r,\delta_r,\mu_f,\delta_f)$ is infinite are both possible, to be compared with Remark 4.5.

6.12 **Theorem** For $0\le\mu_r\le\delta_r,0\le\mu_f\le\delta_f$ arbitrary so that NZC is true, the set

$$D=\{(u,x)\,|\,u\in S,x\in Sol_{RIDC}(u,\mu_r,\delta_r,\mu_f,\delta_f)\} \qquad (1)$$

is a DE.

**Proof** The satisfaction of 3.8 i) is a consequence of 6.11 and the satisfaction of 3.8 ii) coincides with the inclusion 6.5 (1) and we show 3.8 iii). Let $d \in \mathbf{R}$ arbitrary. The translation of 6.2 ii) with $d$ is

$$\overline{(x\circ\tau^d)(t-0)}\cdot(x\circ\tau^d)(t)=\overline{x(t-d-0)}\cdot x(t-d)\leq \bigcap_{\xi\in[t-d-\delta_r,t-d-\delta_r+\mu_r]}u(\xi) =$$

$$= \bigcap_{\xi+d\in[t-\delta_r,t-\delta_r+\mu_r]}u(\xi) = \bigcap_{\xi\in[t-\delta_r,t-\delta_r+\mu_r]}(u\circ\tau^d)(\xi) \qquad (2)$$

together with the dual statement. The hypothesis states that $u\circ\tau^d \in S$, thus (see 2.23) we have that $\bigcap_{\xi\in[\cdot-\delta_r,\cdot-\delta_r+\mu_r]}(u\circ\tau^d)(\xi)\in S$ and we infer that $\min\ supp \bigcap_{\xi\in[\cdot-\delta_r,\cdot-\delta_r+\mu_r]}(u\circ\tau^d)(\xi)\geq 0$. From (2) we get

$$\min\ supp\ x\circ\tau^d = \min\ supp\ \overline{(x\circ\tau^d)(\cdot-0)}\cdot(x\circ\tau^d)(\cdot)\geq$$

$$\geq \min\ supp \bigcap_{\xi\in[\cdot-\delta_r,\cdot-\delta_r+\mu_r]}(u\circ\tau^d)(\xi)\geq 0 \qquad (3)$$

implying $x\circ\tau^d \in S$ from 2.19 b). (2) and its dual shows that $(u\circ\tau^d, x\circ\tau^d)\in D$. 3.8 iii) is proved.

6.13 **Definition** Any DE $D'\subset D$, where the set $D$ is defined by 6.12 (1) is called the (*bivalent*) *relative inertial delay element* (RIDE) (or *circuit* or *buffer*).

6.14 **Theorem** Let RIDC described by 6.2 i), ii) where $0\leq\mu_r\leq\delta_r, 0\leq\mu_f\leq\delta_f$ and NZC is satisfied. The next property is true: $\forall u, \exists x\in Sol_{RIDC}(u,\mu_r,\delta_r,\mu_f,\delta_f), \forall t,$

$$\overline{x(t-0)}\cdot x(t)=\overline{x(t-0)}\cdot \bigcap_{\xi\in[t-\delta_r,t-\delta_r+\mu_r]}u(\xi) \qquad (1)$$

$$x(t-0)\cdot\overline{x(t)}=x(t-0)\cdot \bigcap_{\xi\in[t-\delta_f,t-\delta_f+\mu_f]}\overline{u(\xi)} \qquad (2)$$

The solution of the system (1), (2) is unique.

**Proof** We define

$$x(t)=\begin{cases}1, & \bigcap_{\xi\in[t-\delta_r,t-\delta_r+\mu_r]} u(\xi)=1\\ 0, & \bigcap_{\xi\in[t-\delta_f,t-\delta_f+\mu_f]} \overline{u(\xi)}=1\\ x(t-0), else\end{cases} \quad (3)$$

and, due to NZC –see the analogy from 4.3 b)- this definition is correct. From Lemma 2.23 we know that

- $\bigcap_{\xi\in[t-\delta_r,t-\delta_r+\mu_r]} u(\xi)$ is a signal
- $\bigcap_{\xi\in[t-\delta_f,t-\delta_f+\mu_f]} \overline{u(\xi)} = \overline{\bigcup_{\xi\in[t-\delta_f,t-\delta_f+\mu_f]} u(\xi)}$ is the complement of a signal, satisfying the properties 2.12 b.1), 2.12 b.3) but not 2.12 b.2) and the conclusion is that $x$ given by (3) satisfies $supp\ x\subset[0,\infty)$ and on the other hand that it has limits and only left discontinuities, given by the left discontinuities of $\bigcap_{\xi\in[t-\delta_r,t-\delta_r+\mu_r]} u(\xi)$ , $\bigcap_{\xi\in[t-\delta_f,t-\delta_f+\mu_f]} \overline{u(\xi)}$ :

$$\overline{x(t-0)}\cdot x(t)\le \overline{u(t-\delta_r-0)}\cdot \bigcap_{\xi\in[t-\delta_r,t-\delta_r+\mu_r]} u(\xi) \quad (4)$$

$$x(t-0)\cdot \overline{x(t)}\le u(t-\delta_f-0)\cdot \bigcap_{\xi\in[t-\delta_f,t-\delta_f+\mu_f]} \overline{u(\xi)} \quad (5)$$

in other words $x$ is a signal. Moreover, $x$ satisfies (1), (2) in both situations

- if $\bigcap_{\xi\in[t-\delta_r,t-\delta_r+\mu_r]} u(\xi)$ , $\bigcap_{\xi\in[t-\delta_f,t-\delta_f+\mu_f]} \overline{u(\xi)}$ are null and in this case (1), (2) have all the four terms null, from (4), (5)
- if one of $\bigcap_{\xi\in[t-\delta_r,t-\delta_r+\mu_r]} u(\xi)$ , $\bigcap_{\xi\in[t-\delta_f,t-\delta_f+\mu_f]} \overline{u(\xi)}$ is 1 and the other is null, from (3).

We have proved that (3) implies (1), (2). The equivalence of these statements will be proved at 8.1, by following a different line of demonstration.

We prove the uniqueness of the solution (3) and let us suppose that (1), (2) has two distinct solutions $x, x'$ that necessarily satisfy the property that $t' \geq 0$ exists so that

$$\forall t < t', x(t) = x'(t) \tag{6}$$

$$x(t') = 0, x'(t') = 1 \tag{7}$$

(1), (2) become at the time instant $t'$, taking into account (7):

$$0 = \overline{x(t'-0)} \cdot \bigcap_{\xi \in [t'-\delta_r, t'-\delta_r+\mu_r]} u(\xi) \tag{8}$$

$$x(t'-0) = x(t'-0) \cdot \bigcap_{\xi \in [t'-\delta_f, t'-\delta_f+\mu_f]} \overline{u(\xi)} \tag{9}$$

$$\overline{x'(t'-0)} = \overline{x'(t'-0)} \cdot \bigcap_{\xi \in [t'-\delta_r, t'-\delta_r+\mu_r]} u(\xi) \tag{10}$$

$$0 = x'(t'-0) \cdot \bigcap_{\xi \in [t'-\delta_f, t'-\delta_f+\mu_f]} \overline{u(\xi)} \tag{11}$$

and on the other hand (6) implies

$$x(t'-0) = x'(t'-0) \tag{12}$$

(8), (10), (12) and respectively (9), (11), (12) give

$$x'(t'-0) = 1 \tag{13}$$

$$x(t'-0) = 0 \tag{14}$$

(12), (13) and (14) are contradictory and this ends the proof of the fact that the solution of (1), (2) is unique.

6.15 **Counterexample** If $D, D'$ are RIDE's, then their serial connection $D'' = D' \circ D$ is not in general an RIDE. We give in this sense the counterexample from Fig 9, for which the inequations are given by (1),…,(4):

$$\overline{x(t-0)} \cdot x(t) \leq \overline{x(t-0)} \cdot \bigcap_{\xi \in [t-2, t-1]} u(\xi) \tag{1}$$

$$x(t-0) \cdot \overline{x(t)} \leq x(t-0) \cdot \bigcap_{\xi \in [t-2, t-1]} \overline{u(\xi)} \tag{2}$$

$$\overline{y(t-0)} \cdot y(t) \leq \overline{y(t-0)} \cdot \bigcap_{\xi \in [t-3, t-1]} x(\xi) \tag{3}$$

$$y(t-0) \cdot \overline{y(t)} \leq y(t-0) \cdot \bigcap_{\xi \in [t-3, t-1]} \overline{x(\xi)} \tag{4}$$

and from the rising-falling symmetry conditions, the serial connection should satisfy the inequations:

$$\overline{y(t-0)}\cdot y(t)\leq \overline{y(t-0)}\cdot \bigcap_{\xi\in[t-\delta,t-\delta+\mu]} u(\xi) \tag{5}$$

$$y(t-0)\cdot \overline{y(t)}\leq y(t-0)\cdot \bigcap_{\xi\in[t-\delta,t-\delta+\mu]} \overline{u(\xi)} \tag{6}$$

where $0\leq\mu\leq\delta$ are parameters to be identified from the request that (1),…,(6) are all satisfied as equalities. Such solutions $x\in Sol_{RIDC}(u,1,2,1,2)$, $y\in Sol_{RIDC}(x,2,3,2,3)$ always exist in a unique manner by 6.14 since NZC is fulfilled twice and if $D''=D'\circ D$ is a RIDE, then such a solution $y'\in Sol_{RIDC}(u,\mu,\delta,\mu,\delta)$ should always exist since NZC is fulfilled and moreover we should have $y'=y$.

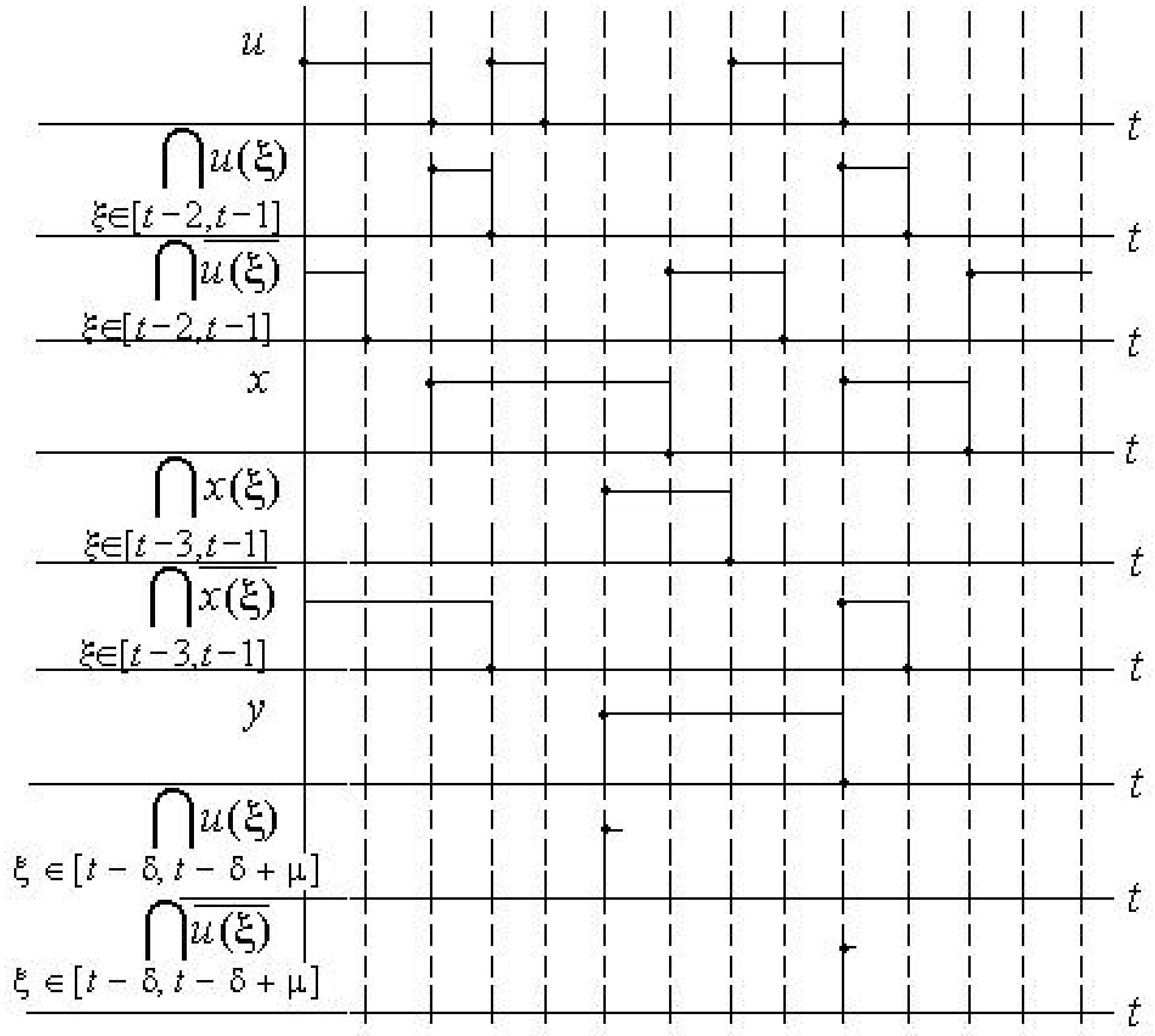

Fig 9

We have

$$t=5:\quad \overline{y(t-0)}\cdot y(t)=1 \Rightarrow \bigcap_{\xi\in[t-\delta,t-\delta+\mu]} u(\xi)=1 \; and \; \delta=5 \; and \; 2>\mu \qquad (7)$$

$$t=9:\quad y(t-0)\cdot \overline{y(t)}=1 \Rightarrow \bigcap_{\xi\in[t-\delta,t-\delta+\mu]} \overline{u(\xi)}=1 \; and \; 3>\mu \qquad (8)$$

$$t=12:\quad \overline{y(t-0)}\cdot y(t)=0 \Rightarrow \bigcap_{\xi\in[t-\delta,t-\delta+\mu]} u(\xi)=0 \; and \; 2\le\mu \qquad (9)$$

But (7) and (9) are contradictory, thus (5) and (6) are not true. This has occurred because the pulse $u(\xi)=1, \xi\in[0,2)$ has produced $\overline{y(5-0)}\cdot y(5)=1$, but the 'similar' pulse $u(\xi)=1, \xi\in[7,9)$ has not produced $\overline{y(12-0)}\cdot y(12)=1$. The source of the situation consists in the fact that the system (1), (2) identifies $u(\xi)=\begin{cases}1, \xi\in[0,2)\vee[3,4)\\ 0, \xi\in[2,3)\end{cases}$ with $u(\xi)=1, \xi\in[0,4)$ and this results by looking in Fig 9 at the form of $x$.

6.16 **Example** of RIDC: $m_r=m_f=0$ and, NZC being satisfied, $d_f\le d_r$,

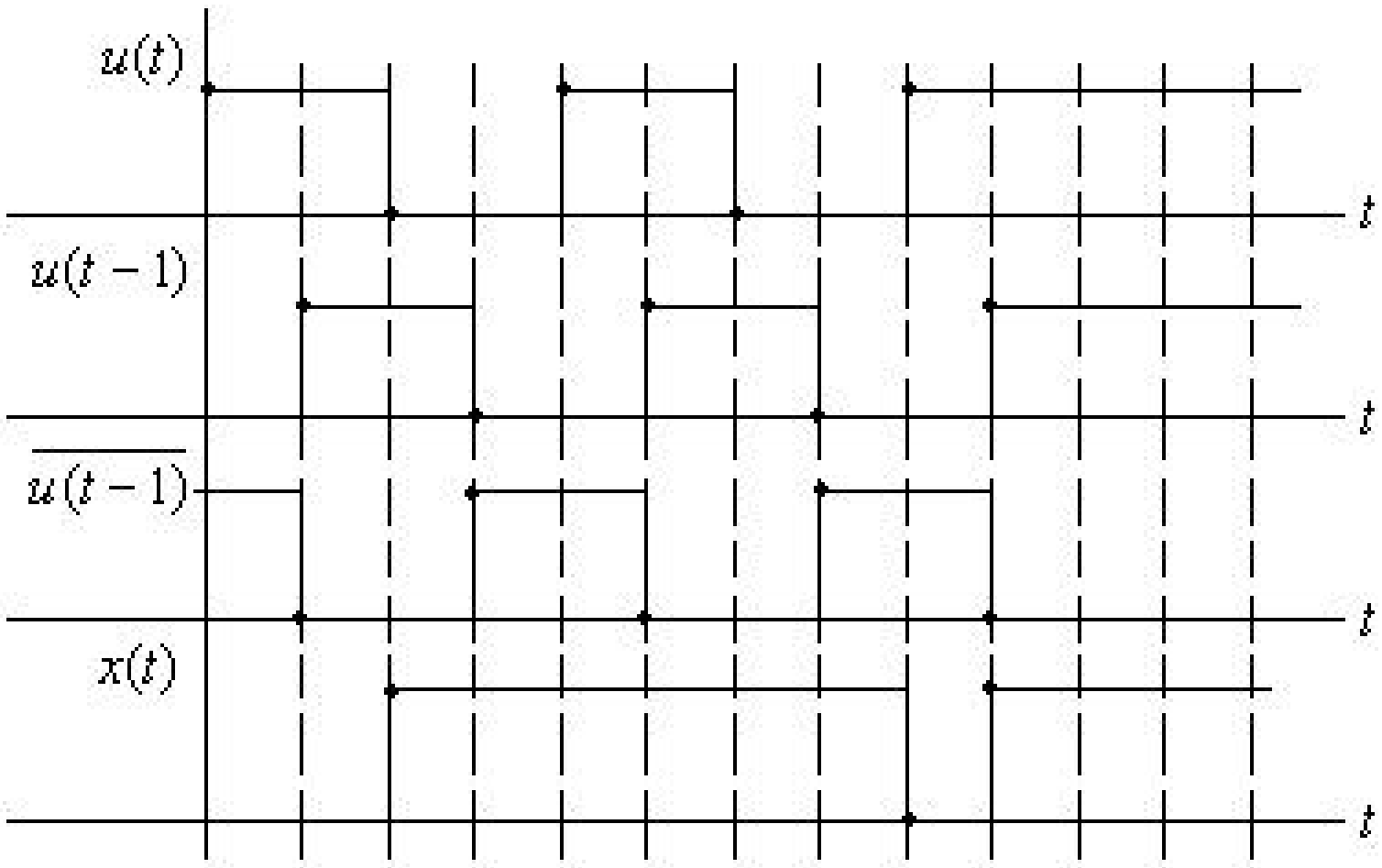

Fig 10

$d_r\le d_f$, thus $d_r=d_f=d$. 6.1 (1) and 6.1 (3) are written under the form:

$$\overline{x(t-0)}\cdot x(t)\le u(t-d) \qquad (1)$$

$$x(t-0)\cdot\overline{x(t)}\leq\overline{u(t-d)} \qquad (2)$$

Taking into account SC also, we have the situation from Fig 10, where we have supposed that $d=1$.

**7. Bounded Bivalent Relative Inertial Delays**

7.1 **Definition** Let $u,x\in S$ and the real numbers $0\leq m_r\leq d_r$, $0\leq m_f\leq d_f$, $0\leq\mu_r\leq\delta_r$, $0\leq\mu_f\leq\delta_f$. The next system of inequations:

i) $$\bigcap_{\xi\in[t-d_r,t-d_r-m_r]}u(\xi)\leq x(t)\leq\bigcup_{\xi\in[t-d_f,t-d_f+m_f]}u(\xi) \qquad (1)$$

ii) $$\overline{x(t-0)}\cdot x(t)\leq\bigcap_{\xi\in[t-\delta_r,t-\delta_r+\mu_r]}u(\xi) \qquad (2)$$

$$x(t-0)\cdot\overline{x(t)}\leq\bigcap_{\xi\in[t-\delta_f,t-\delta_f+\mu_f]}\overline{u(\xi)} \qquad (3)$$

is called the *bounded* (*bivalent*) *relative inertial delay condition* (BRIDC); $u,x,m_r,d_r,m_f,d_f,\mu_r,\delta_r,\mu_f,\delta_f$ have the previous names We say that the tuple $(u,m_r,d_r,m_f,d_f,\mu_r,\delta_r,\mu_f,\delta_f,x)$ satisfies BRIDC. A RIDC that is not bounded is called *unbounded*.

7.2 **Notation** We note with

$$Sol_{BRIDC}(u,m_r,d_r,m_f,d_f,\mu_r,\delta_r,\mu_f,\delta_f)=$$
$$=\{x\,|\,(u,m_r,d_r,m_f,d_f,\mu_r,\delta_r,\mu_f,\delta_f,x)\ \textit{satisfies BRIDC}\} \qquad (1)$$

the set of the solutions of BRIDC.

7.3 **Remark** The meaning of BRIDC results from Fig 11.

7.4 **Theorem** BRIDC has solutions for any $u$ if and only if one of the next requests is satisfied:

a) $$d_f-m_f\leq\delta_r\leq d_r\leq\delta_r-\mu_r+m_r$$
$$d_r-m_r\leq\delta_f\leq d_f\leq\delta_f-\mu_f+m_f$$

b) $$d_r-m_r+\mu_r\leq\delta_r\leq d_f-m_f\leq d_r$$
$$d_f-m_f+\mu_f\leq\delta_f\leq d_r-m_r\leq d_f$$

c) $$d_f-m_f\leq\delta_r\leq d_r-m_r+\mu_r\leq d_r$$

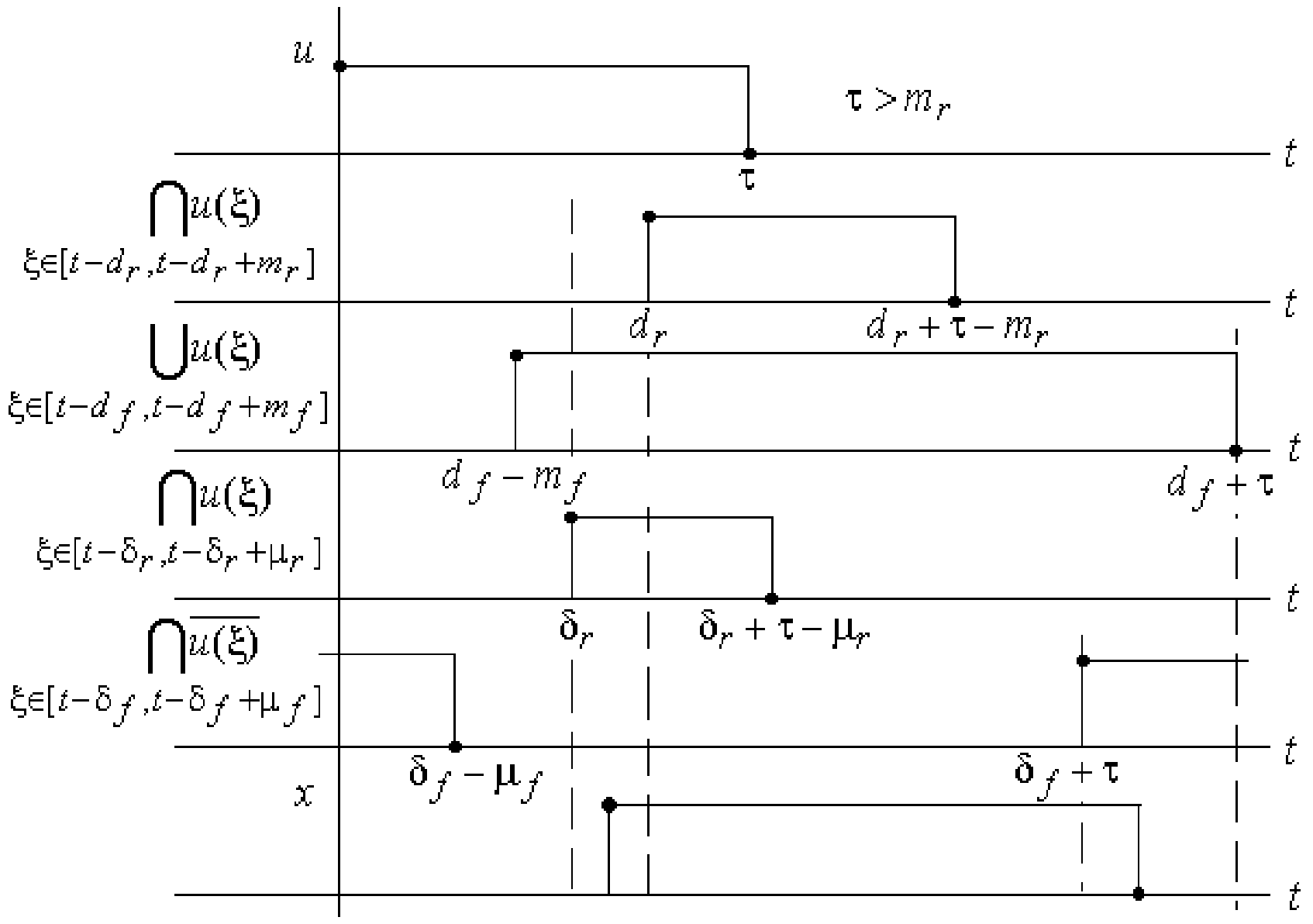

Fig 11

d) $$d_r - m_r \le \delta_f \le d_f - m_f + \mu_f \le d_f$$
$$\delta_r \le d_f - m_f \le \delta_r + m_r - \mu_r \le d_r$$
$$\delta_f \le d_r - m_r \le \delta_f + m_f - \mu_f \le d_f$$

**Proof** Solutions exist iff whenever $x$ must have the value 1, respectively the value 0 in $t$ ( $\bigcap\limits_{\xi\in[t-d_r,t-d_r+m_r]} u(\xi)$ switches in $t$ from 0 to 1, respectively $\bigcup\limits_{\xi\in[t-d_f,t-d_f+m_f]} u(\xi)$ switches in $t$ from 1 to 0), RIDC gives this possibility ($t'$ exists so that $\bigcap\limits_{\xi\in[t'-\delta_r,t'-\delta_r+\mu_r]} u(\xi)=1$, respectively so that $\bigcap\limits_{\xi\in[t'-\delta_f,t'-\delta_f+\mu_f]} \overline{u(\xi)}=1$) in time ($t'\in[t-d_r+d_f-m_f,t]$, respectively $t'\in[t-d_f+d_r-m_r,t]$).

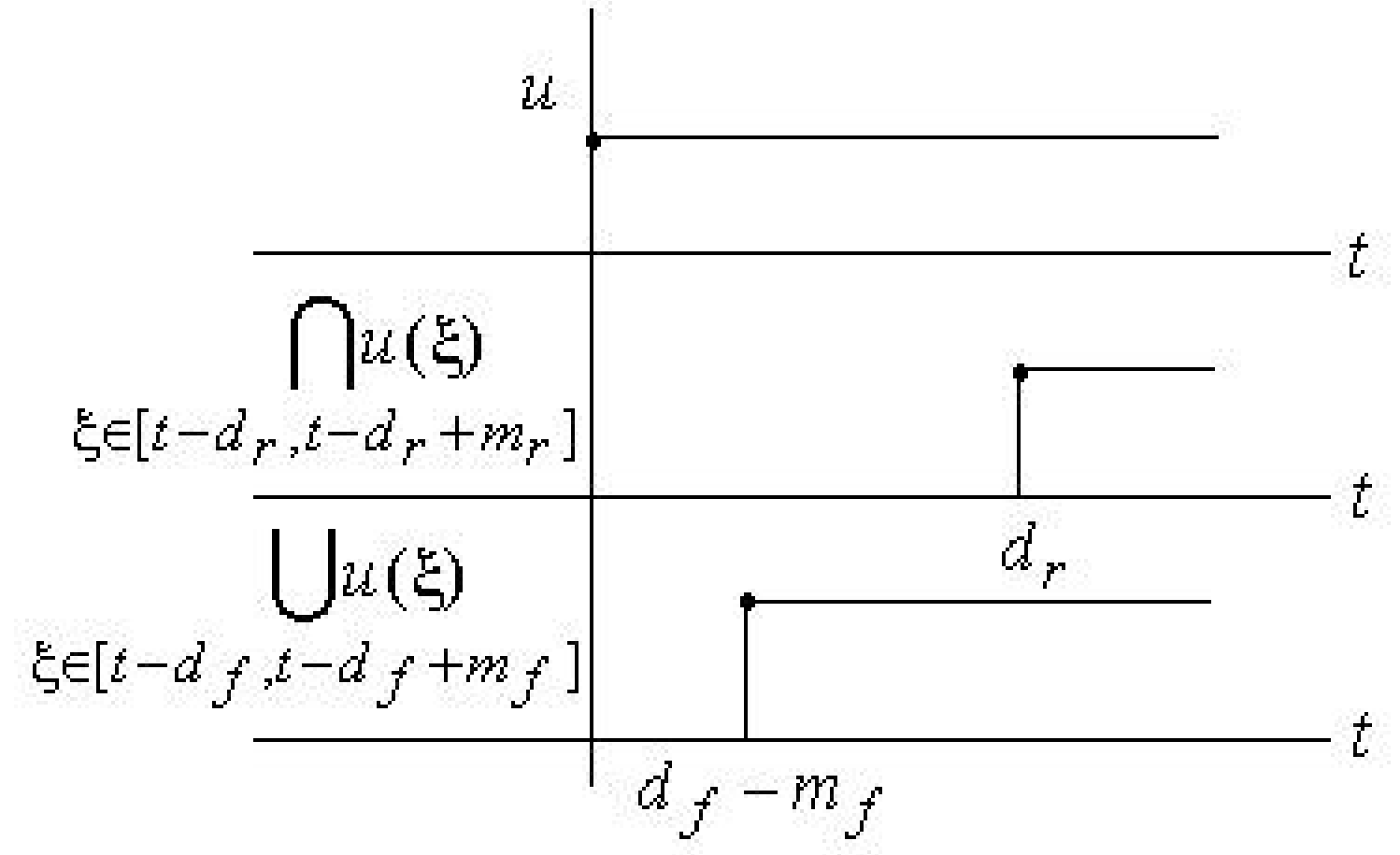

Fig 12

This happens for example in Fig 12 for $t = d_r$. We can write

$$\overline{u(t-d_r-0)}\cdot \bigcap_{\xi\in[t-d_r,t-d_r+m_r]} u(\xi) \le \bigcup_{t'\in[t-d_r+d_f-m_f,t]} \bigcap_{\xi\in[t'-\delta_r,t'-\delta_r+\mu_r]} u(\xi)$$

inequality that is true for any $u$. The next statements are all equivalent with the previous one:

$$\exists t'\in[t-d_r+d_f-m_f,t],[t-d_r,t-d_r+m_r]\supset[t'-\delta_r,t'-\delta_r+\mu_r]$$

$$\exists t', \begin{cases} t-d_r+d_f-m_f \le t' \\ t-d_r+\delta_r \le t' \end{cases} and \begin{cases} t' \le t \\ t' \le t-d_r+m_r+\delta_r-\mu_r \end{cases}$$

$$\max(t-d_r+d_f-m_f, t-d_r+\delta_r) \le \min(t, t-d_r+m_r+\delta_r-\mu_r)$$

one of the next possibilities is true:

j) $-d_r+d_f-m_f \le -d_r+\delta_r$

$0 \le -d_r+m_r+\delta_r-\mu_r$

$-d_r+\delta_r \le 0$

jj) $-d_r+d_f-m_f \ge -d_r+\delta_r$

$0 \le -d_r+m_r+\delta_r-\mu_r$

$-d_r+d_f-m_f \le 0$

jjj) $-d_r+d_f-m_f \le -d_r+\delta_r$

$$0 \geq -d_r + m_r + \delta_r - \mu_r$$

$$-d_r + \delta_r \leq -d_r + m_r + \delta_r - \mu_r$$

jv) $$-d_r + d_f - m_f \geq -d_r + \delta_r$$

$$0 \geq -d_r + m_r + \delta_r - \mu_r$$

$$-d_r + d_f - m_f \leq -d_r + m_r + \delta_r - \mu_r$$

It is shown that j), jj), jjj), jv) are equivalent with the first statements of a), b), c), d).

7.5 **Definition** The condition

7.4 a) *or* 7.4 b) *or* 7.4 c) *or* 7.4 d)

is called the *consistency condition* (CC) associated with BRIDC 7.1.

7.6 **Example** We suppose that the following conditions, stronger than CC (they imply 7.4 a)) are fulfilled:

$$d_f - m_f \leq \delta_f - \mu_f \leq \delta_r \leq d_r \qquad (1)$$

$$d_r - m_r \leq \delta_r - \mu_r \leq \delta_f \leq d_f \qquad (2)$$

In this case, BRIDC with $u$ having sufficiently long pulses ($u$ is 1 strictly longer than $m_r$ and then $u$ is 0 strictly longer than $m_f$) is characterized by time intervals of the following kind:

- $t \in (-\infty, d_f - m_f)$; $x(t) = 0$ and the only possible switch -that does not happen- is from 1 to 0
- $t \in [d_f - m_f, \delta_f - \mu_f)$; $x(t) = 0$. $x(t)$ could be 1, but it is allowed to switch only from 1 to 0 and this does not happen
- $t \in [\delta_f - \mu_f, \delta_r)$; $x(t) = 0$. $x(t)$ could be 1, but no switch is allowed
- $t \in [\delta_r, d_r]$; $x(t) = 0$ and $x(t) = 1$ are both allowed, switching from 0 to 1 is allowed to happen and exactly one such switch takes place
- $t \in (d_r, d_r + \tau - m_r)$; $x(t) = 1$ and the only possible switch -that does not happen- is from 0 to 1
- $t \in [d_r + \tau - m_r, \delta_r + \tau - \mu_r)$; $x(t) = 1$, $x(t)$ could be 0, but it is allowed to switch only from 0 to 1 and this does not happen
- $t \in [\delta_r + \tau - \mu_r, \delta_f + \tau)$; $x(t) = 1$, $x(t)$ could be 0, but no switches are possible in this time interval
- $t \in [\delta_f + \tau, d_f + \tau]$; $x(t) = 1$ and $x(t) = 0$ are both possible,

switching from 1 to 0 is allowed and exactly one such switch takes place

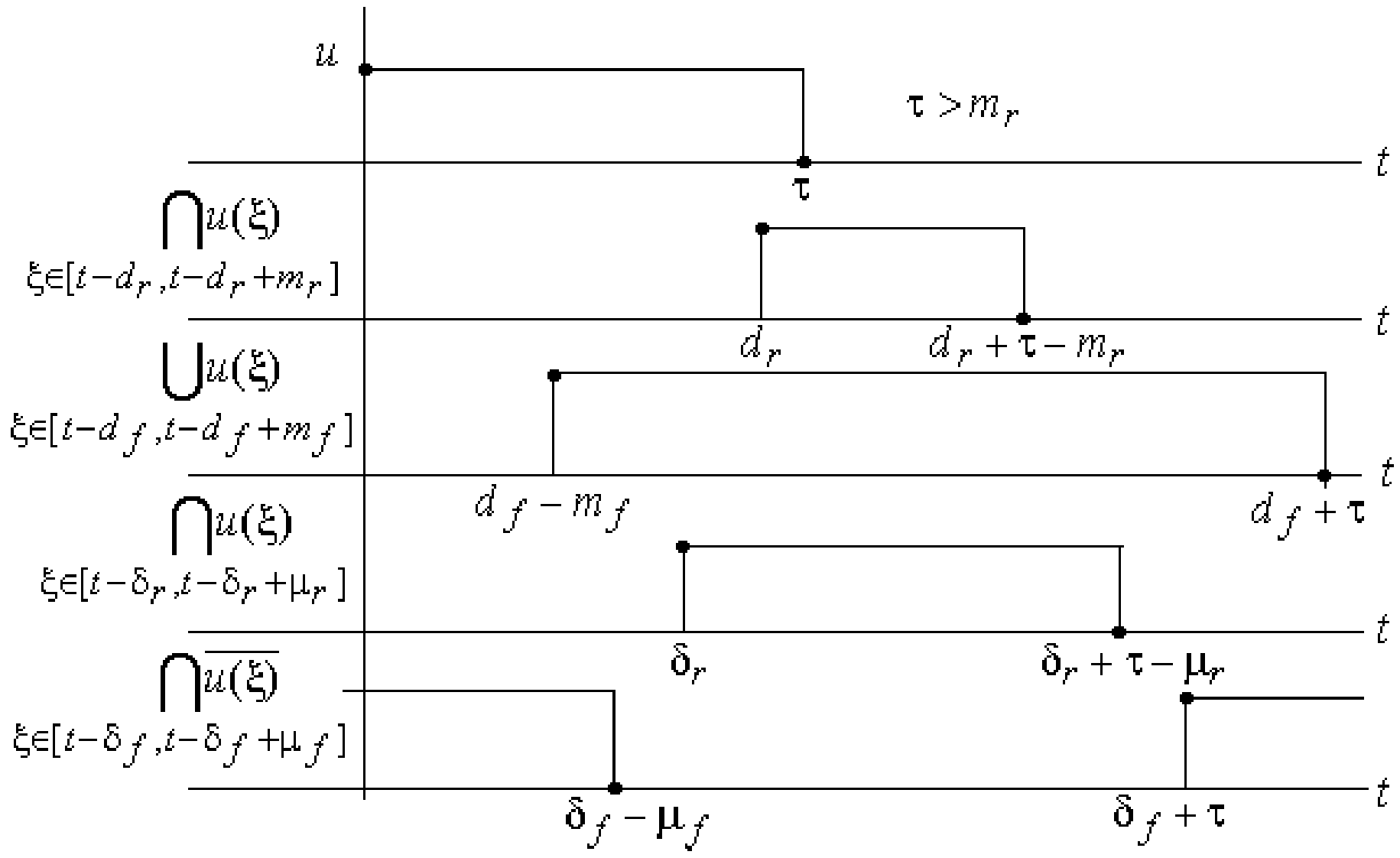

Fig 13

- $t \in (d_f + \tau, \infty)$; $x(t) = 0$ and the only possible switch -that does not happen- is from 1 to 0.

This succession of time intervals is repetitive if $u$ is suitable chosen.

7.7 **Theorem** (*Special case of BRIDC*) When $d_r = \delta_r, d_f = \delta_f$ and $m_r = \mu_r, m_f = \mu_f$ and if $d_f \geq d_r - m_r, d_r \geq d_f - m_f$ is true, then the next systems:

$$\bigcap_{\xi\in[t-d_r, t-d_r-m_r]} u(\xi) \leq x(t) \leq \bigcup_{\xi\in[t-d_f, t-d_f+m_f]} u(\xi) \qquad (1)$$

$$\overline{x(t-0)} \cdot x(t) \leq \bigcap_{\xi\in[t-d_r, t-d_r+m_r]} u(\xi) \qquad (2)$$

$$x(t-0) \cdot \overline{x(t)} \leq \bigcap_{\xi\in[t-d_f, t-d_f+m_f]} \overline{u(\xi)} \qquad (3)$$

and respectively

$$\overline{x(t-0)}\cdot x(t)=\overline{x(t-0)}\cdot\bigcap_{\xi\in[t-d_r,t-d_r+m_r]}u(\xi)\qquad(4)$$

$$x(t-0)\cdot\overline{x(t)}=x(t-0)\cdot\bigcap_{\xi\in[t-d_f,t-d_f+m_f]}\overline{u(\xi)}\qquad(5)$$

are equivalent, in the sense that the signals $u,x$ satisfy the first system if and only if they satisfy the second system.

**Proof** By multiplying both the left inequality of (1) and the inequality (2) with $\overline{x(t-0)}$, we have:

$$\overline{x(t-0)}\cdot x(t)\leq\overline{x(t-0)}\cdot\bigcap_{\xi\in[t-d_r,t-d_r+m_r]}u(\xi)\leq\overline{x(t-0)}\cdot x(t-0)\qquad(6)$$

from where we get (4).

The complementation of the right inequality of (1) gives

$$\bigcap_{\xi\in[t-d_f,t-d_f+m_f]}\overline{u(\xi)}\leq\overline{x(t)}\qquad(7)$$

By the multiplication of (3) and (7) with $x(t-0)$ we have

$$x(t-0)\cdot\overline{x(t)}\leq x(t-0)\cdot\bigcap_{\xi\in[t-d_f,t-d_f+m_f]}\overline{u(\xi)}\leq x(t-0)\cdot\overline{x(t)}\qquad(8)$$

thus we have obtained (5).

Conversely, (4) implies (2) and (5) implies (3). In order to show that the system (4), (5) implies (1), we suppose that $\bigcap_{\xi\in[t-d_r,t-d_r+m_r]}u(\xi)=1$ for some arbitrary $t$. If $\overline{x(t-0)}=1$, then $x(t)=1$ from (4) thus the left inequality of (1) is proved and if $\overline{x(t-0)}=0$, then (5) becomes $\overline{x(t)}=0$, thus $x(t)=1$ and the left inequality of (1) is proved again. Proving the fact that (4), (5) imply the validity of the right inequation of (1) is similar.

7.8 **Remark** Let $0\leq m_r\leq d_r, 0\leq m_f\leq d_f, 0\leq\mu_r\leq\delta_r, 0\leq\mu_f\leq\delta_f$ be arbitrary. We have $Sol_{BRIDC}(\mathbf{0},m_r,d_r,m_f,d_f,\mu_r,\delta_r,\mu_f,\delta_f)=\{\mathbf{0}\}$; $u\neq\mathbf{0}$ and CC imply that $Sol_{BRIDC}(u,m_r,d_r,m_f,d_f,\mu_r,\delta_r,\mu_f,\delta_f)=\{\mathbf{0}\}$ and $Sol_{BRIDC}(u,m_r,d_r,m_f,d_f,\mu_r,\delta_r,\mu_f,\delta_f)$ is infinite are both possible.

7.9 **Remark** From the manner in which BRIDC was defined we have, if CC is satisfied:

$$Sol_{BRIDC}(u,m_r,d_r,m_f,d_f,\mu_r,\delta_r,\mu_f,\delta_f)=$$
$$=Sol_{BDC}(u,m_r,d_r,m_f,d_f)\wedge Sol_{RIDC}(u,\mu_r,\delta_r,\mu_f,\delta_f) \quad (1)$$

7.10 **Theorem** For $0\leq m_r\leq d_r, 0\leq m_f\leq d_f, 0\leq \mu_r\leq \delta_r, 0\leq \mu_f\leq \delta_f$ arbitrary so that CC is fulfilled, the set

$$D=\{(u,x)\,|\,u\in S, x\in Sol_{BRIDC}(u,m_r,d_r,m_f,d_f,\mu_r,\delta_r,\mu_f,\delta_f)\} \quad (1)$$

is a DE.

**Proof** The satisfaction of 3.8 a) results from 7.8, the satisfaction of 3.8 b) from 7.9 because

$$Sol_{BRIDC}(u,m_r,d_r,m_f,d_f,\mu_r,\delta_r,\mu_f,\delta_f)\subset$$
$$\subset Sol_{BDC}(u,m_r,d_r,m_f,d_f)\subset Sol_{SC}(u)$$

and the validity of 3.8 c) may be shown by combining the similar properties for BDE's and RIDE's.

7.11 **Definition** Any DE $D'\subset D$, where the set $D$ was defined at 7.10 (1) is called the *bounded* (*bivalent*) *relative inertial delay element* (BRIDE) (*circuit*, *buffer*). $D$ itself is called the *full* BRIDE. A RIDE that is not bounded is called *unbounded*.

7.12 Let the BRIDE's $D, D'$ and their serial connection $D''=D'\circ D$. $D''$ is not a BRIDE in general, but a BDE due to the fact that the serial connection of the RIDE's is not a RIDE.

## 8. Deterministic Bivalent Relative Inertial Delays

8.1 **Theorem** Let the real numbers $0\leq m_r\leq d_r, 0\leq m_f\leq d_f$ arbitrary with $d_r-m_r\leq d_f$, $d_f-m_f\leq d_r$. The next systems are equivalent, in the sense that if $u,x\in S$ satisfy one of them, then they also satisfy any other.

a)
$$\overline{x(t-0)}\cdot x(t)=\overline{x(t-0)}\cdot \bigcap_{\xi\in[t-d_r,t-d_r+m_r]} u(\xi) \quad (1)$$
$$x(t-0)\cdot \overline{x(t)}=x(t-0)\cdot \bigcap_{\xi\in[t-d_f,t-d_f+m_f]} \overline{u(\xi)} \quad (2)$$

b) $$\bigcap_{\xi\in[t-d_r,t-d_r+m_r]} u(\xi) \le x(t) \tag{3}$$

$$\bigcap_{\xi\in[t-d_f,t-d_f+m_f]} \overline{u(\xi)} \le \overline{x(t)} \tag{4}$$

$$\overline{\bigcap_{\xi\in[t-d_r,t-d_r+m_r]} u(\xi)} \cdot \overline{\bigcap_{\xi\in[t-d_f,t-d_f+m_f]} \overline{u(\xi)}} \le$$

$$\le \overline{x(t-0)} \cdot \overline{x(t)} \cup x(t-0) \cdot x(t) \tag{5}$$

c) $$x(t) = \begin{cases} 1, & \bigcap_{\xi\in[t-d_r,t-d_r+m_r]} u(\xi) = 1 \\ 0, & \bigcap_{\xi\in[t-d_f,t-d_f+m_f]} \overline{u(\xi)} = 1 \\ x(t-0), & otherwise \end{cases} \tag{6}$$

d) $$x(t) = \bigcap_{\xi\in[t-d_r,t-d_r+m_r]} u(\xi) \cup x(t-0) \cdot \bigcup_{\xi\in[t-d_f,t-d_f+m_f]} u(\xi) \tag{7}$$

e) $$Dx(t) = \overline{x(t-0)} \cdot \bigcap_{\xi\in[t-d_r,t-d_r+m_r]} u(\xi) \cup$$

$$\cup\, x(t-0) \cdot \bigcap_{\xi\in[t-d_f,t-d_f+m_f]} \overline{u(\xi)} \tag{8}$$

f) $$\overline{x(t-0)} \cdot x(t) \cdot \bigcap_{\xi\in[t-d_r,t-d_r+m_r]} u(\xi) \cup x(t-0) \cdot \overline{x(t)} \cdot \bigcap_{\xi\in[t-d_f,t-d_f+m_f]} \overline{u(\xi)} \cup$$

$$\cup\, \overline{x(t-0)} \cdot \overline{x(t)} \cdot \overline{\bigcap_{\xi\in[t-d_r,t-d_r+m_r]} u(\xi)} \cup$$

$$\cup\, x(t-0) \cdot x(t) \cdot \overline{\bigcap_{\xi\in[t-d_f,t-d_f+m_f]} \overline{u(\xi)}} = 1 \tag{9}$$

**Proof** We shall show the implications $a) \Rightarrow b) \Rightarrow c) \Rightarrow d) \Rightarrow e) \Rightarrow a)$ and the equivalence $a) \Leftrightarrow f)$ (some of the previous implications prove to be equivalences during the proof)

$a) \Rightarrow b)$ If $\bigcap_{\xi\in[t-d_r,t-d_r+m_r]} u(\xi) = 1$, then $\bigcap_{\xi\in[t-d_f,t-d_f+m_f]} \overline{u(\xi)} = 0$ from CC and $x(t) = 1$ is the unique solution of the system (1), (2):

$$\overline{x(t-0)}\cdot x(t)=\overline{x(t-0)} \quad (10)$$

$$x(t-0)\cdot\overline{x(t)}=0 \quad (11)$$

(3) is proved.

(4) may be proved in a dual manner.

The supposition $\bigcap_{\xi\in[t-d_r,t-d_r+m_r]} u(\xi) = \bigcap_{\xi\in[t-d_f,t-d_f+m_f]} \overline{u(\xi)} = 0$ gives the conclusion that $\overline{x(t-0)}\cdot x(t)\cup x(t-0)\cdot\overline{x(t)}=0$, from where $\overline{\overline{x(t-0)}\cdot x(t)\cup x(t-0)\cdot\overline{x(t)}}=\overline{x(t-0)}\cdot\overline{x(t)}\cup x(t-0)\cdot x(t)=1$. (5) is proved.

$b)\Rightarrow c)$ If $\bigcap_{\xi\in[t-d_r,t-d_r+m_r]} u(\xi) = 1$, then $x(t)=1$ and if $\bigcap_{\xi\in[t-d_f,t-d_f+m_f]} \overline{u(\xi)} = 1$, then $\overline{x(t)}=1$ and thus $x(t)=0$. Otherwise, $\overline{x(t-0)}\cdot\overline{x(t)}\cup x(t-0)\cdot x(t)=1$, i.e. $x(t)=x(t-0)$.

$c)\Rightarrow d)$ Because $\bigcap_{\xi\in[t-d_r,t-d_r+m_r]} u(\xi) \le \bigcup_{\xi\in[t-d_f,t-d_f+m_f]} u(\xi)$, we have the three possibilities 4.10 (2), 4.10 (3), 4.10 (4).

Case i), 4.10 (2) is true. Then $\overline{\bigcup_{\xi\in[t-d_f,t-d_f+m_f]} u(\xi)} = \bigcap_{\xi\in[t-d_f,t-d_f+m_f]} \overline{u(\xi)} = 1$ and $x(t)=0$. (7) represents the equation $x(t)=0$ too.

Case ii), 4.10 (3) is true. Then $\bigcap_{\xi\in[t-d_f,t-d_f+m_f]} \overline{u(\xi)} = 0$ and (6), (7) coincide both with the equation $x(t)=x(t-0)$.

Case iii), 4.10 (4) is true. In this situation (6), (7) become both $x(t)=1$.

$d)\Rightarrow e)$ We have the same three possibilities.

Case i), 4.10 (2) is true and d) means that $x(t)=0$. Because $\bigcap_{\xi\in[t-d_f,t-d_f+m_f]} \overline{u(\xi)} = 1$, e) gives $x(t)\oplus x(t-0)=x(t-0)$ thus $x(t)=0$ and (7), (8) coincide.

Case ii) 4.10 (3) is true and d) means that $x(t)=x(t-0)$, in other words

$Dx(t)=0$. On the other hand $\overline{\bigcap_{\xi\in[t-d_f,t-d_f+m_f]} u(\xi)}=0$ and e) becomes $Dx(t)=0$ too.

Case iii) 4.10 (4) is true and d) means that $x(t)=1$. Because $\bigcap_{\xi\in[t-d_f,t-d_f+m_f]} \overline{u(\xi)}=0$, e) implies $x(t)\oplus x(t-0)=\overline{x(t-0)}=$
$=x(t-0)\oplus 1$, i.e. $x(t)=1$.

$e)\Rightarrow a)$ We make use once again of the three cases i), ii), iii).

Case i), 4.10 (2) is satisfied. Because $\overline{\bigcup_{\xi\in[t-d_f,t-d_f+m_f]} u(\xi)}=$
$=\bigcap_{\xi\in[t-d_f,t-d_f+m_f]} \overline{u(\xi)}=1$, e) means $Dx(t)=x(t-0)$ i.e. $x(t)=0$; on the other hand, the equations (1), (2) become

$$\overline{x(t-0)}\cdot x(t)=0 \qquad (12)$$

$$x(t-0)\cdot \overline{x(t)}=x(t-0) \qquad (13)$$

The system (12), (13) is uniquely satisfied by $x(t)=0$.

Case ii), 4.10 (3) is true. e) shows that $Dx(t)=0$ and a) gives $\overline{x(t-0)}\cdot x(t)=x(t-0)\cdot\overline{x(t)}=0$ i.e. $\overline{x(t-0)}\cdot x(t)\cup x(t-0)\cdot\overline{x(t)}=0=$
$=Dx(t)$.

Case iii), 4.10 (4) is satisfied. e) gives $Dx(t)=\overline{x(t-0)}$, i.e. $x(t-0)\oplus x(t)=x(t-0)\oplus 1$ and $x(t)=1$. a) is equivalent with the system (10), (11) whose unique solution is $x(t)=1$.

$a)\Leftrightarrow f)$ We observe that a) is equivalent with the system

$$\overline{x(t-0)}\cdot x(t)\cdot\overline{x(t-0)}\cdot\bigcap_{\xi\in[t-d_r,t-d_r+m_r]} u(\xi)\ \cup$$
$$\cup\ \overline{\overline{x(t-0)}\cdot x(t)}\cdot\overline{\overline{x(t-0)}\cdot\bigcap_{\xi\in[t-d_r,t-d_r+m_r]} u(\xi)}=1$$

$$x(t-0)\cdot\overline{x(t)}\cdot x(t-0)\cdot\bigcap_{\xi\in[t-d_f,t-d_f+m_f]} \overline{u(\xi)}\ \cup$$

$$\cup\, \overline{x(t-0)\cdot \overline{x(t)}}\cdot \overline{x(t-0)\cdot \bigcap_{\xi\in[t-d_f,\,t-d_f+m_f]} \overline{u(\xi)}} = 1$$

thus with the equations

$$1 = (\overline{x(t-0)}\cdot x(t)\cdot \bigcap_{\xi\in[t-d_r,\,t-d_r+m_r]} u(\xi)\ \cup$$

$$\cup\, (x(t-0)\cup \overline{x(t)})\cdot (x(t-0)\cup \overline{\bigcap_{\xi\in[t-d_r,\,t-d_r+m_r]} u(\xi)}\ ))\cdot$$

$$\cdot (x(t-0)\cdot \overline{x(t)}\cdot \bigcap_{\xi\in[t-d_f,\,t-d_f+m_f]} \overline{u(\xi)}\ \cup$$

$$\cup\, (\overline{x(t-0)}\cup x(t))\cdot (\overline{x(t-0)}\cup \overline{\bigcap_{\xi\in[t-d_f,\,t-d_f+m_f]} \overline{u(\xi)}}\ )) =$$

$$= (\overline{x(t-0)}\cdot x(t)\cdot \bigcap_{\xi\in[t-d_r,\,t-d_r+m_r]} u(\xi)\ \cup$$

$$\cup\, (x(t-0)\cup \overline{x(t)})\cdot \overline{\bigcap_{\xi\in[t-d_r,\,t-d_r+m_r]} u(\xi)}\ \cup x(t-0))\cdot$$

$$\cdot (x(t-0)\cdot \overline{x(t)}\cdot \bigcap_{\xi\in[t-d_f,\,t-d_f+m_f]} \overline{u(\xi)}\ \cup$$

$$\cup\, (\overline{x(t-0)}\cup x(t))\cdot \overline{\bigcap_{\xi\in[t-d_f,\,t-d_f+m_f]} \overline{u(\xi)}}\ \cup \overline{x(t-0)}) =$$

$$= x(t-0)\cdot \overline{x(t)}\cdot \overline{\bigcap_{\xi\in[t-d_r,\,t-d_r+m_r]} u(\xi)}\cdot \bigcap_{\xi\in[t-d_f,\,t-d_f+m_f]} \overline{u(\xi)}\ \cup$$

$$\cup\, x(t-0)\cdot \overline{x(t)}\cdot \bigcap_{\xi\in[t-d_f,\,t-d_f+m_f]} \overline{u(\xi)}\ \cup$$

$$\cup\, \overline{x(t-0)}\cdot x(t)\cdot \bigcap_{\xi\in[t-d_r,\,t-d_r+m_r]} u(\xi)\cdot \overline{\bigcap_{\xi\in[t-d_f,\,t-d_f+m_f]} \overline{u(\xi)}}\ \cup$$

$$\cup\, (\overline{x(t-0)}\cdot \overline{x(t)}\cup x(t-0)\cdot x(t))\cdot \overline{\bigcap_{\xi\in[t-d_r,\,t-d_r+m_r]} u(\xi)}\cdot \overline{\bigcap_{\xi\in[t-d_f,\,t-d_f+m_f]} \overline{u(\xi)}}$$

$$\cup\, x(t-0)\cdot x(t)\cdot \overline{\bigcap_{\xi\in[t-d_f,\,t-d_f+m_f]} \overline{u(\xi)}}\ \cup$$

$$\cup \overline{x(t-0)} \cdot x(t) \cdot \bigcap_{\xi \in [t-d_r, t-d_r+m_r]} u(\xi) \cup \overline{x(t-0)} \cdot \overline{x(t)} \cdot \overline{\bigcap_{\xi \in [t-d_r, t-d_r+m_r]} u(\xi)}$$

$$= x(t-0) \cdot \overline{x(t)} \cdot \overline{\bigcap_{\xi \in [t-d_r, t-d_r+m_r]} u(\xi)} \cdot \bigcap_{\xi \in [t-d_f, t-d_f+m_f]} \overline{u(\xi)} \cup$$

$$\cup x(t-0) \cdot \overline{x(t)} \cdot \bigcap_{\xi \in [t-d_f, t-d_f+m_f]} \overline{u(\xi)} \cup$$

$$\cup \overline{x(t-0)} \cdot x(t) \cdot \bigcap_{\xi \in [t-d_r, t-d_r+m_r]} u(\xi) \cdot \overline{\bigcap_{\xi \in [t-d_f, t-d_f+m_f]} \overline{u(\xi)}} \cup$$

$$\cup \overline{x(t-0)} \cdot x(t) \cdot \bigcap_{\xi \in [t-d_r, t-d_r+m_r]} u(\xi) \cup$$

$$\cup \overline{x(t-0)} \cdot \overline{x(t)} \cdot \overline{\bigcap_{\xi \in [t-d_r, t-d_r+m_r]} u(\xi)} \cdot \overline{\bigcap_{\xi \in [t-d_f, t-d_f+m_f]} \overline{u(\xi)}} \cup$$

$$\cup \overline{x(t-0)} \cdot \overline{x(t)} \cdot \overline{\bigcap_{\xi \in [t-d_r, t-d_r+m_r]} u(\xi)} \cup$$

$$\cup x(t-0) \cdot x(t) \cdot \overline{\bigcap_{\xi \in [t-d_r, t-d_r+m_r]} u(\xi)} \cdot \overline{\bigcap_{\xi \in [t-d_f, t-d_f+m_f]} \overline{u(\xi)}} \cup$$

$$\cup x(t-0) \cdot x(t) \cdot \overline{\bigcap_{\xi \in [t-d_f, t-d_f+m_f]} \overline{u(\xi)}}$$

$$= x(t-0) \cdot \overline{x(t)} \cdot \bigcap_{\xi \in [t-d_f, t-d_f+m_f]} \overline{u(\xi)} \cup \overline{x(t-0)} \cdot x(t) \cdot \bigcap_{\xi \in [t-d_r, t-d_r+m_r]} u(\xi) \cup$$

$$\cup \overline{x(t-0)} \cdot \overline{x(t)} \cdot \overline{\bigcap_{\xi \in [t-d_r, t-d_r+m_r]} u(\xi)} \cup x(t-0) \cdot x(t) \cdot \overline{\bigcap_{\xi \in [t-d_f, t-d_f+m_f]} \overline{u(\xi)}}$$

8.2 **Remark** The equations 8.1 a) have occurred for the first time at 6.14 and at 6.15 they were used as a counterexample showing that the serial connection of the RIDE's is not in general an RIDE. We have also met them at 7.7 as a special case of BRIDC, where $d_r = \delta_r, d_f = \delta_f, m_r = \mu_r$, $m_f = \mu_f$ are true.

8.3 **Remark** In any of the equivalent conditions 8.1 a),…,8.1 f), CC from 7.4 coincides with $d_r \geq d_f - m_f$, $d_f \geq d_r - m_r$. We shall refer to it as the *consistency condition* CC too.

8.4 **Remark** The implications of the possible violation of CC in one of 8.1 a),...,8.1 f) are the following. Let $u$ and $t'$ so that

$$[t'-d_r, t'-d_r+m_r] \wedge [t'-d_f, t'-d_f+m_f] = \emptyset \tag{1}$$

$$\forall \xi \in [t'-d_r, t'-d_r+m_r], u(\xi) = 1 \tag{2}$$

$$\forall \xi \in [t'-d_f, t'-d_f+m_f], u(\xi) = 0 \tag{3}$$

Let us suppose, in order to make a choice, that $t'-d_r+m_r < t'-d_f$. From (2), (3) and from the right continuity of $u$ in $t'-d_r+m_r, t'-d_f+m_f$ we get the existence of $\delta > 0$ so that $t'-d_r+m_r+\delta < t'-d_f$ and

$$\forall \xi \in [t'-d_r, t'-d_r+m_r+\delta], u(\xi) = 1 \tag{4}$$

$$\forall \xi \in [t'-d_f, t'-d_f+m_f+\delta], u(\xi) = 0 \tag{5}$$

8.1 a) becomes for any $t \in [t', t'+\delta]$

$$\overline{x(t-0)} \cdot x(t) = \overline{x(t-0)} \tag{6}$$

$$x(t-0) \cdot \overline{x(t)} = x(t-0) \tag{7}$$

and the system accepts two possibilities $x(t) = 0, x(t-0) = 1$ and $x(t) = 1$, $x(t-0) = 0$; no signal $x$ satisfies such requests. 8.1 b) has no solution either, because 8.1 (3), 8.1 (4) show that $x(t) = 0$ and $x(t) = 1$ are both true for $t \in [t', t'+\delta]$ and this is the case of 8.1 c) also, where $x(t)$ is not a well defined function for $t \in [t', t'+\delta]$. 8.1 d) gives $x(t) = 1$, $t \in [t', t'+\delta]$. 8.1 e) and 8.1 f) become both

$$Dx(t) = 1, t \in [t', t'+\delta] \tag{8}$$

and this equation represents a nonsense, because the set $\{t \mid t \in [t', t'+\delta], Dx(t) = 1\}$ should be finite ($x$ has resulted to be 'everywhere discontinuous' in $[t', t'+\delta]$).

8.5 **Theorem** We suppose for the numbers $0 \le m_r \le d_r, 0 \le m_f \le d_f$ that CC is satisfied. Then any of 8.1 a),...,8.1 f) has a solution, that is unique.
**Proof** This follows from 8.1 and 6.14.

8.6 **Definition** For $u, x \in S$ and $0 \le m_r \le d_r, 0 \le m_f \le d_f$ so that the consistency condition is satisfied, any of the equivalent properties 8.1 a), ..., 8.1 f) and respectively 7.7 (1), (2), (3) is called the *deterministic* (*bivalent*) *relative inertial delay condition* (DRIDC). We say that the tuple $(u, m_r, d_r, m_f, d_f, x)$ satisfies DRIDC.

8.7 **Remark** We give an interpretation for the statements from 8.1.

a) $x$ was 0 and $u$ was 1 for sufficiently long iff $x$ switches from 0 to 1 + the dual statement

b) if $u$ was 1 for sufficiently long, then $x$ is 1 + the dual statement; if $u$ was not sufficiently persistent in order that it is reproduced at the output, then $x$ keeps its previous value

c), d) similar with b)

e) the switches of $x$ occur exactly when either $x$ was 0 and $u$ was 1 for sufficiently long, or $x$ was 1 and $u$ was 0 for sufficiently long

f) at any moment in time we have one of the next situations true:

- $x$ switches from 0 to 1 and $u$ was 1 for sufficiently long + the dual statement

- $x$ keeps the 0 value and $u$ was not 1 for sufficiently long + the dual statement

8.8 **Notation** We suppose that $0 \le m_r \le d_r, 0 \le m_f \le d_f$ and that CC is satisfied. The set consisting in the unique solution $x$ (see Theorem 8.5) of DRIDC is noted with

$$Sol_{DRIDC}(u, m_r, d_r, m_f, d_f) =$$
$$= \{x \mid (u, m_r, d_r, m_f, d_f, x) \text{ satisfies DRIDC}\} \quad (1)$$

8.9 **Theorem** For all $u$ and all $m_r, d_r, m_f, d_f$ like at 8.8, we have

$$Sol_{DRIDC}(u, m_r, d_r, m_f, d_f) =$$
$$= Sol_{BRIDC}(u, m_r, d_r, m_f, d_f, m_r, d_r, m_f, d_f) \quad (1)$$

**Proof** This is the result expressed by 7.7.

8.10 **Remark** We have the special case of DRIDC when $m_r = m_f = 0$, $d_r = d_f = d$:

$$\overline{x(t-0)} \cdot x(t) = \overline{x(t-0)} \cdot u(t-d) \quad (1)$$

$$x(t-0) \cdot \overline{x(t)} = x(t-0) \cdot \overline{u(t-d)} \quad (2)$$

The solution of (1), (2) is unique and it is:

$$x(t) = u(t-d) \quad (3)$$

i.e. we have obtained FDC 5.1 (1).

8.11 **Theorem** Let $0 \le m_r \le d_r, 0 \le m_f \le d_f$ arbitrary with $d_r - m_r \le d_f$, $d_f - m_f \le d_r$. The set

$$D = \{(u,x) \mid u \in S, x \in Sol_{DRIDC}(u, m_r, d_r, m_f, d_f)\} \qquad (1)$$

is a deterministic DE.

**Proof** This follows from 7.10, taking into account 8.9.

8.12 **Definition** $D$ previously defined is called the *deterministic* (*bivalent*) *relative inertial delay element* (DRIDE) (or *circuit* or *buffer*).

8.13 **Theorem** Let the DRIDE $D$ and the FDE $I_d$, where $d \ge 0$. Then

$$D' = D \circ I_d = I_d \circ D = \{(u, x \circ \tau^d) \mid (u,x) \in D\} \qquad (1)$$

is a DRIDE with $d_r' = d_r + d, m_r' = m_r, d_f' = d_f + d, m_f' = m_f$.

**Proof** Formula (1) takes place for any DE's D, $I_d$ like in Theorem 5.11. The equations are:

$$x(t) = \bigcap_{\xi \in [t-d_r, t-d_r+m_r]} u(\xi) \cup x(t-0) \cdot \bigcup_{\xi \in [t-d_f, t-d_f+m_f]} u(\xi) \qquad (2)$$

$$y(t) = x(t-d) \qquad (3)$$

$$y(t) = \bigcap_{\xi \in [t-d_r-d, t-d_r-d+m_r]} u(\xi) \cup y(t-0) \cdot \bigcup_{\xi \in [t-d_f-d, t-d_f-d+m_f]} u(\xi) \qquad (4)$$

8.14 **Remark** The serial connection of the DRIDE's is a BDE, but not a DRIDE, as known from 6.15, for example.

## 9. A Comparison with Other Works of the Author

9.1 **Remark** Because differences exist between the points of view from this paper and other points of view that have been expressed in previous works, the purpose of this chapter is to make a comparison between them.

9.2 **Lemma** For $0 \le m_r \le d_r, 0 \le m_f \le d_f$ and $u \in S$ the next formulas are true:

$$\bigcap_{\xi \in [t-d_r, t-d_r+m_r]} u(\xi) = u(t-d_r+m_r) \cdot \overline{\bigcup_{\xi \in (t-d_r, t-d_r+m_r]} Du(\xi)} \qquad (1)$$

$$\bigcap_{\xi \in [t-d_f, t-d_f+m_f]} \overline{u(\xi)} = \overline{u(t-d_f+m_f)} \cdot \overline{\bigcup_{\xi \in (t-d_f, t-d_f+m_f]} Du(\xi)} \qquad (2)$$

**Proof** (1), (2) are trivial if $m_r$, respectively $m_f$ are null and the reunions are null too. If $m_r > 0$, the equality (1) is a consequence of the fact that

$$\bigcap_{\xi\in[t-d_r,t-d_r+m_r]} u(\xi) = 1 \Leftrightarrow u(t-d_r+m_r)=1 \textit{ and } u_{|[t-d_r,t-d_r+m_r]} \textit{ is constant}$$

$$\Leftrightarrow u(t-d_r+m_r)=1 \textit{ and } u_{|(t-d_r,t-d_r+m_r]} \textit{ is constant}$$

(from the right continuity of $u$ in $t-d_r$)

$$\Leftrightarrow u(t-d_r+m_r)=1 \textit{ and } Du_{|(t-d_r,t-d_r+m_r]} \textit{ is null}$$

Such equivalencies, that are easy to accept intuitively, follow from [13] 6.4 Theorem 1 however.

9.3 **Example** We rewrite 8.1 (8) in the special case when

$$d_r = d_f = d \qquad (1)$$

$$m_r = m_f = m > 0 \qquad (2)$$

and we apply Lemma 9.2:

$$Dx(t) = \overline{x(t-0)} \cdot \bigcap_{\xi\in[t-d,t-d+m]} u(\xi) \cup x(t-0) \cdot \bigcap_{\xi\in[t-d,t-d+m]} \overline{u(\xi)} \qquad (3)$$

$$= (\overline{x(t-0)} \cdot u(t-d+m) \cup x(t-0) \cdot \overline{u(t-d+m)}) \cdot \overline{\bigcup_{\xi\in(t-d,t-d+m]} Du(\xi)}$$

In [12], the equation of the inertial delay circuit with null initial conditions was written under the form

$$Dx(t) = (x(t-0) \oplus u(t-0)) \cdot \overline{\bigcup_{\xi\in(t-d,t)} Du(\xi)} \qquad (4)$$

By comparing (3), (4) we reach the conclusion that they are 'equivalent' if we make $m < d$ infinitely close to $d$. This fact corresponds to the ideas exposed in [1] and elsewhere showing that the cancellation delay $m$ and the transmission delay for transitions $d$ are usually taken to be equal.

9.4 **Theorem** The next properties are equivalent in the sense that the arbitrary signals $u, x$ satisfy one of them if and only if they satisfy the other one.

a) $m_r, d_r, m_f, d_f, \mu_r, \delta_r, \mu_f, \delta_f$ are given and the next inequalities are fulfilled (see also Example 7.6)

$$0 \le m_r \le d_r,\ 0 \le m_f \le d_f \qquad (1)$$

$$0 \le \mu_r \le \delta_r, 0 \le \mu_f \le \delta_f \quad (2)$$

$$d_f - m_f \le \delta_f - \mu_f \le \delta_r \le d_r \quad (3)$$

$$d_r - m_r \le \delta_r - \mu_r \le \delta_f \le d_f \quad (4)$$

$$\bigcap_{\xi \in [t-d_r, t-d_r+m_r]} u(\xi) \le x(t) \le \bigcup_{\xi \in [t-d_f, t-d_f+m_f]} u(\xi) \quad (5)$$

$$\overline{x(t-0)} \cdot x(t) \le \bigcap_{\xi \in [t-\delta_r, t-\delta_r+\mu_r]} u(\xi) \quad (6)$$

$$x(t-0) \cdot \overline{x(t)} \le \bigcap_{\xi \in [t-\delta_f, t-\delta_f+\mu_f]} \overline{u(\xi)} \quad (7)$$

b) The numbers $m_{r,\min}, d_{r,\min}, m_{r,\max}, d_{r,\max}, m_{f,\min}, d_{f,\min}, m_{f,\max}, d_{f,\max}$ are given and we have

$$0 \le m_{r,\max} \le d_{r,\max}, 0 \le m_{f,\max} \le d_{f,\max} \quad (8)$$

$$0 \le m_{r,\min} \le d_{r,\min}, 0 \le m_{f,\min} \le d_{f,\min} \quad (9)$$

$$d_{f,\max} - m_{f,\max} \le d_{f,\min} - m_{f,\min} \le d_{r,\min} \le d_{r,\max} \quad (10)$$

$$d_{r,\max} - m_{r,\max} \le d_{r,\min} - m_{r,\min} \le d_{f,\min} \le d_{f,\max} \quad (11)$$

$$\overline{x(t-0)} \cdot \bigcap_{\xi \in [t-d_{r,\max}, t-d_{r,\max}+m_{r,\max}]} u(\xi) \le \overline{x(t-0)} \cdot x(t) \le \quad (12)$$

$$\le \overline{x(t-0)} \cdot \bigcap_{\xi \in [t-d_{r,\min}, t-d_{r,\min}+m_{r,\min}]} u(\xi)$$

$$x(t-0) \cdot \bigcap_{\xi \in [t-d_{f,\max}, t-d_{f,\max}+m_{f,\max}]} \overline{u(\xi)} \le x(t-0) \cdot \overline{x(t)} \le \quad (13)$$

$$\le x(t-0) \cdot \bigcap_{\xi \in [t-d_{f,\min}, t-d_{f,\min}+m_{f,\min}]} \overline{u(\xi)}$$

**Proof** The next equalities take place

$$m_{r,\min} = \mu_r, d_{r,\min} = \delta_r \quad (14)$$

$$m_{r,\max} = m_r, d_{r,\max} = d_r \quad (15)$$

$$m_{f,\min} = \mu_f, d_{f,\min} = \delta_f \quad (16)$$

$$m_{f,\max} = m_f, d_{f,\max} = d_f \quad (17)$$

under the form 'equal by definition with' in both directions $a) \Rightarrow b)$ and $b) \Rightarrow a)$ resulting that (1),…,(4) and (8),…,(11) coincide.

$a) \Rightarrow b)$ The left inequality of (5) multiplied with $\overline{x(t-0)}$ gives the left inequality of (12) and (6) multiplied with $\overline{x(t-0)}$ gives the right inequality of (12). The rest results by duality.

$b) \Rightarrow a)$ We suppose that $\bigcap_{\xi \in [t-d_r, t-d_r+m_r]} u(\xi) = 1$ and we have the next possibilities

i) $x(t-0) = 0$

Then the left inequality of (12) shows that $x(t) = 1$ and the left inequality of (5) is satisfied.

ii) $x(t-0) = 1$

and the right inequality of (13) becomes

$$\overline{x(t)} \leq \bigcap_{\xi \in [t-\delta_f, t-\delta_f+\mu_f]} \overline{u(\xi)} \tag{18}$$

(10), (11) are written under the form (3), (4) and this implies

$$t - \delta_f + \mu_f \geq t - d_r \tag{19}$$

$$t - d_r + m_r \geq t - \delta_f \tag{20}$$

i.e.

$$[t - d_r, t - d_r + m_r] \wedge [t - \delta_f, t - \delta_f + \mu_f] \neq \emptyset \tag{21}$$

and thus $\bigcap_{\xi \in [t-\delta_f, t-\delta_f+\mu_f]} \overline{u(\xi)} = 0$. From (18) we get $x(t) = 1$ and the left inequality of (5) is satisfied in this case too.

On the other hand, the right inequality of (12) gives

$$\overline{x(t-0)} \cdot x(t) \leq \overline{x(t-0)} \cdot \bigcap_{\xi \in [t-\delta_r, t-\delta_r+\mu_r]} u(\xi) \leq \bigcap_{\xi \in [t-\delta_r, t-\delta_r+\mu_r]} u(\xi) \tag{22}$$

i.e. (6).

The other implications result by duality.

9.5 **Remark** We compare 9.4 b) with the inequations

$$\overline{x(t-0)} \cdot \bigcap_{\xi \in [t-d_{r,\max}, t)} u(\xi) \leq \overline{x(t-0)} \cdot x(t) \leq \overline{x(t-0)} \cdot \bigcap_{\xi \in [t-d_{r,\min}, t)} u(\xi) \tag{1}$$

$$x(t-0) \cdot \bigcap_{\xi \in [t-d_{f,\max}, t)} \overline{u(\xi)} \leq x(t-0) \cdot \overline{x(t)} \leq x(t-0) \cdot \bigcap_{\xi \in [t-d_{f,\min}, t)} \overline{u(\xi)} \tag{2}$$

from [11]. We suppose that for some sufficiently small $\varepsilon > 0$ and for $0 < d_{r,\min} \le d_{r,\max}$ , $0 < d_{f,\min} \le d_{f,\max}$ given, we have:

$$m_{r,\max} = d_{r,\max} - \varepsilon, m_{f,\max} = d_{f,\max} - \varepsilon \qquad (3)$$

$$m_{r,\min} = d_{r,\min} - \varepsilon, m_{f,\min} = d_{f,\min} - \varepsilon \qquad (4)$$

When $\varepsilon \to 0$, the two systems of inequations are 'equivalent'.

## 10 Conclusions

The central idea of the theory is represented by SC, DC and the Definition 3.8 of the DE's. They are strengthened sometimes under the form BDC, BDE and interfere with (bivalent) inertia and occasionally with determinism. It is interesting and important to observe the manner in which the serial connection of the DE's preserves the properties of boundness, inertia, CC, NZC and determinism because certain 'failures' occur, caused by inertia.